\documentclass[longauth]{aa}  

\newcommand{\gC}{$\gamma$\,Cas\xspace}
\usepackage{graphicx}
\usepackage{subcaption}

\usepackage{txfonts}

\usepackage{natbib}
\bibpunct{(}{)}{;}{a}{}{,}
\usepackage[colorlinks=true,linkcolor=blue,citecolor=blue, urlcolor=black]{hyperref}

\begin{document} 
	
	\title{Short-term variability and mass loss in Be stars}
	\subtitle{V. Space photometry and ground-based spectroscopy of $\gamma$\,Cas\thanks{Based on data collected by the BRITE-Constellation satellite mission, designed, built, launched, operated and supported by the Austrian Research Promotion Agency (FFG), the University of Vienna, the Technical University of Graz, the University of Innsbruck, the Canadian Space Agency (CSA), the University of Toronto Institute for Aerospace Studies (UTIAS), the Foundation for Polish Science \& Technology (FNiTP MNiSW), and National Science Centre (NCN).
} 
	}
\titlerunning{Photometry and Spectroscopy of $\gamma$\,Cas}

	\author{C. C. Borre  
		\inst{1,2} 
		\and
		D. Baade\inst{2} 
		\and
		A. Pigulski\inst{3} 
		\and
		D. Panoglou\inst{4} 
		\and
		A. Weiss\inst{5} 
		\and
		Th. Rivinius\inst{6} 
		\and
		G. Handler \inst{7} 
		\and
		A. F. J. Moffat\inst{8} 
	    \and
	    \\
		A. Popowicz\inst{9}
		\and
		G. A. Wade\inst{10}
		\and
		W. W. Weiss\inst{11}
		\and 
		K. Zwintz\inst{12}
		}
	
	\institute{Stellar Astrophysics Centre, Department of Physics and Astronomy, Aarhus University, Ny Munkegade 120, DK-8000 Aarhus C, Denmark; \email{cborre@phys.au.dk}
		\and
		European Organisation for Astronomical Research in the Southern Hemisphere (ESO), Karl-Schwarzschild-Str. 2,
		85748 Garching b. München, Germany;		\email{dbaade@eso.org}
		\and
		Instytut Astronomiczny, Uniwersytet Wroc\l awski, ul. Kopernika 11, 51-622 Wroc\l aw, Poland
		\and
		Observat\'orio Nacional, Rua General Jo\'e Cristino 77, S\~ao   Crist\'ov\~ao RJ 20921-400, Rio de Janeiro, Brazil
		\and
		Max-Planck-Institut f\"ur Astrophysik, Karl-Schwarzschild-Str. 1, 85748 Garching, Germany
        \and
		European Organisation for Astronomical Research in the Southern Hemisphere (ESO), Casilla 19001, Santiago 19, Chile
		\and
		Nicolaus Copernicus Astronomical Center, Polish Academy of Sciences,
		ul. Bartycka 18, 00-716, Warsaw, Poland
		\and
        Observatoire Astronomique du Mont Mégantic, Departement de Physique, Université de Montréal C. P. 6128, Succursale: Centre-Ville, Montréal, QC H3C 3J7, Canada
		\and
		Silesian University of Technology, Institute of Automatic Control, Gliwice, Akademicka 16, Poland
        \and 
        Department of Physics \& Space Science, Royal Military College of Canada, PO Box 17000 Station Forces, Kingston, ON K7K 0C6, Canada
        \and
        Institute for Astrophysics, University of Vienna, Tuerkenschanzstrasse 17,
        1180 Vienna, Austria
        \and
		Universit\"at Innsbruck, Institut f\"ur Astro- und Teilchenphysik, Technikerstrasse 25, A-6020 Innsbruck
	}
	
	\date{Received ??; accepted ??}
	

	\abstract
	{Be stars are physically complex systems that continue to challenge theory to understand their rapid rotation, complex variability and decretion disks. $\gamma$\,Cassiopeiae ($\gamma$\,Cas) is one such star but is even more curious because of its unexplained hard thermal X-ray emission.}
	{We aim to examine the optical variability of \gC and thereby to shed more light on its puzzling behaviour.}
	{Three hundred twenty-one archival H$\alpha$ spectra from 2006 to 2017 are analysed to search for frequencies corresponding to the 203.5 day orbit of the companion. Space photometry from the SMEI satellite from 2003 to 2011 and the BRITE-Constellation of nano-satellites between 2015 and 2019 is investigated in the period range from a couple of hours to a few days.}
	{The orbital period of the companion of 203.5 days is confirmed with independent measurements from the structure of the H$\alpha$ line emission. A strong blue/red asymmetry in the amplitude distribution across the H$\alpha$ emission line could hint at a spiral structure in the decretion disk. With the space photometry, the known frequency of 0.82\,d$^{-1}$ is confirmed in data from the early 2000s. A higher frequency of 2.48\,d$^{-1}$ is present in the data from 2015 to 2019 and possibly also in the early 2000s.  A third frequency at 1.25\,d$^{-1}$ is proposed to exist in both SMEI and BRITE data. The only explanation covering all three rapid variations seems to be nonradial pulsation.  The two higher frequencies are incompatible with rotation.
	}
	{}
	
	
	\maketitle
	
	
	\section{Introduction}
\label{sec:intro} 
Classical Be stars are a subgroup of B-type stars with emission lines originating from a Keplerian decretion disk. They were first discovered by \citeauthor{Secchi1866} in 1866 and were named B\textit{e} stars because of the spectral class B and the \textit{e}mission lines. A widely accepted classification for the classical Be stars is: 
\textit{Non-supergiant B stars whose spectra have, or had at some time, one or more Balmer lines in emission} \citep{Jaschek1981,Collins1987}.
For a history of the classification and an overall review on Be stars see \cite{Rivinius_R}.

Be stars are rapid rotators with rotation velocities of $\sim$75\% of the critical rotation and above \citep{Fremat2005,Meilland2012,Rivinius_R}. The origin of the high velocities is still a topic of debate. In binary systems, an explanation is spin-up due to angular momentum transfer from the companion \citep{Kriz&Harmanec1975,Pols1991,deMink2013}. For isolated stars, different mechanisms are proposed such as angular momentum transfer from the core \citep{Ekstrom2008,Granada2013}, past interactions with an undetected companion, or merger events \citep{deMink2011}. It has been estimated that about 1/3 of Be stars are binaries \citep{Abt&Levy1978,Oudmaijer&Perr2010}. However, due to the strong rotational broadening and the frequently high intrinsic brightness and mass of B-type stars, it can be very challenging to detect a companion spectroscopically. 

As discussed by \cite{Rimulo2018} and references therein, Be stars may undergo a surface-angular-momentum crisis as the core contracts and the outer parts spin up. The disk could, through viscosity, provide a way for the angular momentum to be dispersed so that the Be star is not rotationally disrupted \citep[as originally envisioned by][]{Struve1931}.
 
The formation of the disk is also an unsolved problem. It has been suggested that non-radial pulsations (NRPs) in combination with the rapid rotation cause mass ejections into the disk \citep[e.g.][]{Baade1988}. This is supported by the fact that nearly all Be stars display NRPs \citep[see][and references therein]{Rivinius_R}. 
Another suggestion is that tidal interactions with a companion can build the disk \citep{Kriz&Harmanec1975}. However, the rarity of Be stars in binaries implies that such a model would only be able to explain a minority of the Be stars.
Stellar winds were proposed but they are no longer a strong candidate after it was found that the mass-loss rate to the disk is an order of magnitude larger than the winds produced by a B-type star \citep{Carciofi2012,Puls2008}. Furthermore, \citet{Owocki1994} showed that it is not possible to form a Keplerian disk from winds. 
Magnetic fields were also a contender but this hypothesis was rejected due to the lack of field detections in Be-type stars by MiMeS survey \citep{Wade2016}.

After the mass has been ejected, a dust-free Keplerian disk is formed \citep[][to name a few]{Okazaki1991,Papaloizou1992}. The most widely accepted theory for the disk is the viscous decretion disk model \citep[][]{Lee1991,krticka2011}. 
The disk behaves physically as an accretion disk if the time was reversed i.e. the star ejects matter out into the disk instead of accreting it \citep{Balbus2003,Pringle1981,Shakura&Sunyaev1973}. 
As long as the star keeps ejecting matter with appropriate angular momentum the disk is sustained. If the outbursts stop or become less frequent, the disk turns, within months to years, to an accretion disk and the inner parts reaccrete onto the star while the outer parts dissipate \citep{Ghoreyshi2018}. Due to its small size \citep[R$_\mathrm{disk}$$\sim$10 R$_\star$; list in][their Table 2]{Rivinius_R} the disk varies on short timescales and can appear and disappear many times during the lifetime of the Be star. Furthermore, a spiral structure can form in the disk \citep{Kato1983,Okazaki1991,Papaloizou1992}.
 
The H$\alpha$ emission is the strongest optical emission line and its shape and evolution contain much information about the system. \cite{Horne1986} illustrated how different parts of the profile correspond to specific regions in the disk.    
For some binary Be stars, the H$\alpha$ variation caused by a spiral arm has been observed to be phase-locked to the orbital motion of the companion \citep{Stefl2007} while it is not for others.
With simulations, \citet{Panoglou2016} demonstrated how, under suitable circumstances (circular and co-planar orbit with proper radius), the companion can induce or at least influence the spiral structure of the disk. 
The presence of a companion can also create resonances in the disk leading to a truncation as demonstrated by \cite{Okazaki2002}. These authors show that the disk becomes smaller than a disk unperturbed by a companion and that the density may be slightly larger because the disk is "squeezed" by the companion. This can appear as a "flat" top in the emission line structure because the outer parts of the disk - that contribute the most to the peaks at low velocities \citep{Horne1986} - are reduced in size.

The peak structure also depends on the viewing angle i.e. the inclination of the disk. \citet[][their Fig.\,1]{Rivinius_R} presents a sketch on how the structure can change from a "wine bottle" shape when viewed face-on \citep{Hanuschik1986} to an emission line with a narrow absorption core when viewed edge-on. Stars displaying lines with such narrow absorption cores are called shell stars \citep{Hanuschik1995,Rivinius1999}. The shape of the line can change over time see, e.g. \citet{Hanuschik1996} for observations or \citet{Panoglou2018} for simulations on this.

In this paper, we provide an overview of $\gamma$\,Cas and a motivation of this work in Sect.\,\ref{sec:gCas} and a description of the observations in Sect.\,\ref{sec:obs}. The analysis is presented in Sect.\,\ref{sec:analysis} which is split into three parts based on the observing facility. In Sect.\,\ref{sec:discussion} we discuss the findings. A summary and conclusions are given in Sect.\,\ref{sec:conclusion}. 

\section{$\gamma$\,Cas}
\label{sec:gCas}
$\gamma$\,Cas (27 Cas, HR 264, \object{HD 5394}) is a B0.5 IVe star and is arguably the  best known Be star. It was the first of its kind to be discovered by \cite{Secchi1866} and  was deemed the prototype for the class. It was later observed to exhibit hard X-ray emission \citep{Jernigan1976,White1982} which is unusual for early-type stars \citep{Gudel&Naze2009}. Furthermore, the X-ray flux is much lower than in Be-X-ray binaries \citep{Naze2018}. Recent work by \citet{Langer2019} suggests that the X-rays originate from the interaction between the wind from a He-star companion and the Be star disk or wind.
More than a dozen other Be stars have been observed to have similar X-ray emission and this new class of stars is called the \gC analogues or \gC-like stars \citep{Naze2018}. 
 
Using long-baseline optical interferometry, the inclination angle of the star's disk is observed to be around 45$^\circ$ with respect to the line of sight \citep[e.g.][]{Quirrenbach1997}. Like all Be stars, \gC is a rapid rotator and rotates with near-critical velocity \citep{Fremat2005}.   

\gC is a known binary system with an orbital period of 203.52(8) days \citep[e.g.][]{Harmanec2000,Miroshnichenko2002,Nemravova2012}. The orbit is circular or has a very small eccentricity ($e<0.03$). Signatures of the companion have not been observed directly but its presence is inferred from radial velocity measurements of the Be component. The mass of the Be primary is estimated to be 13-16 M$_\odot$ and the secondary between 0.5-1 M$_\odot$ \citep{Harmanec2000,Nemravova2012}. 

\gC went through two shell phases in 1935-36 and 1939-40, both of which were preceded by a large outburst \citep[e.g.][]{Heard1938,Baldwin1940}.
It then appeared for a short time with no emission lines i.e. no disk. In the second half of the 40's the emission in the Balmer lines slowly returned and it was present ever since. 

A few years ago, \citet{Henry2012} \citep[and in the previous work by][]{Smith2006} used the ground-based Automated Photometric Telescope\footnote{\url{http://schwab.tsuniv.edu/t3.html}} (APT), located in southern Arizona, to observe \gC from 1997-2011 in the $V-$ and $B-$band. They found that the time series was dominated by slow 2-3-month variations on which a short periodic variation of 1.21 days (0.82\,d$^{-1}$) was superimposed. This 1.21 day modulation varied in amplitude both in the blue and red wavelength range. During the end of the observation period, it almost disappeared. The period corresponds rather well with their radius and the near-critical rotational velocity estimate and it was accepted as the rotational period of the primary star in \citet{Henry2012}. They did not, however, state which property of the star is modulated by the star's rotation. 

There are also observations of migrating subfeatures moving across photospheric absorption-line profiles \citep[see e.g.][]{Yang1988}. It is still debated whether these features stem from NRPs or from co-rotating and absorbing cloudlets \citep{Smith1998}.
A further overview of the observational history is given by \cite{Harmanec2002}.

The motivation for this new study of \gC is provided by the 
two phenomena that seem to make the supposedly prototypical Be star \gC different from other Be stars, namely the hard X-ray emission and the apparent lack of NRPs.  For an independent assessment of the companion's effect on the Be star \citep[cf.\ ][]{Langer2019} we used the Be Star Spectra database (BeSS) 
to analyse the structural change of the H$\alpha$ line. This database is particularly suitable due to its long time baseline of spectra.
Furthermore, we performed time series analyses of high-cadence space photometry to search for NRPs. For this, an obvious choice of instrument is the BRight Target Explorer (BRITE) because it observes bright stars repeatedly for half-year intervals and with multiple satellites and in two spectral passbands.  A less common source of stellar photometry is the Solar Mass Ejection Imager (SMEI). However, as \citet{Goss2011} have shown for the bright Be star $\alpha$ Eridani, SMEI is very suitable for frequency detection in bright stars. Although the data from SMEI are rather noisy the long time baseline ($\sim$9 years) makes up for the lower quality. In this way, SMEI and BRITE complement each other very well.

		\begin{figure}
			\centering
			\includegraphics[width=8.8cm]{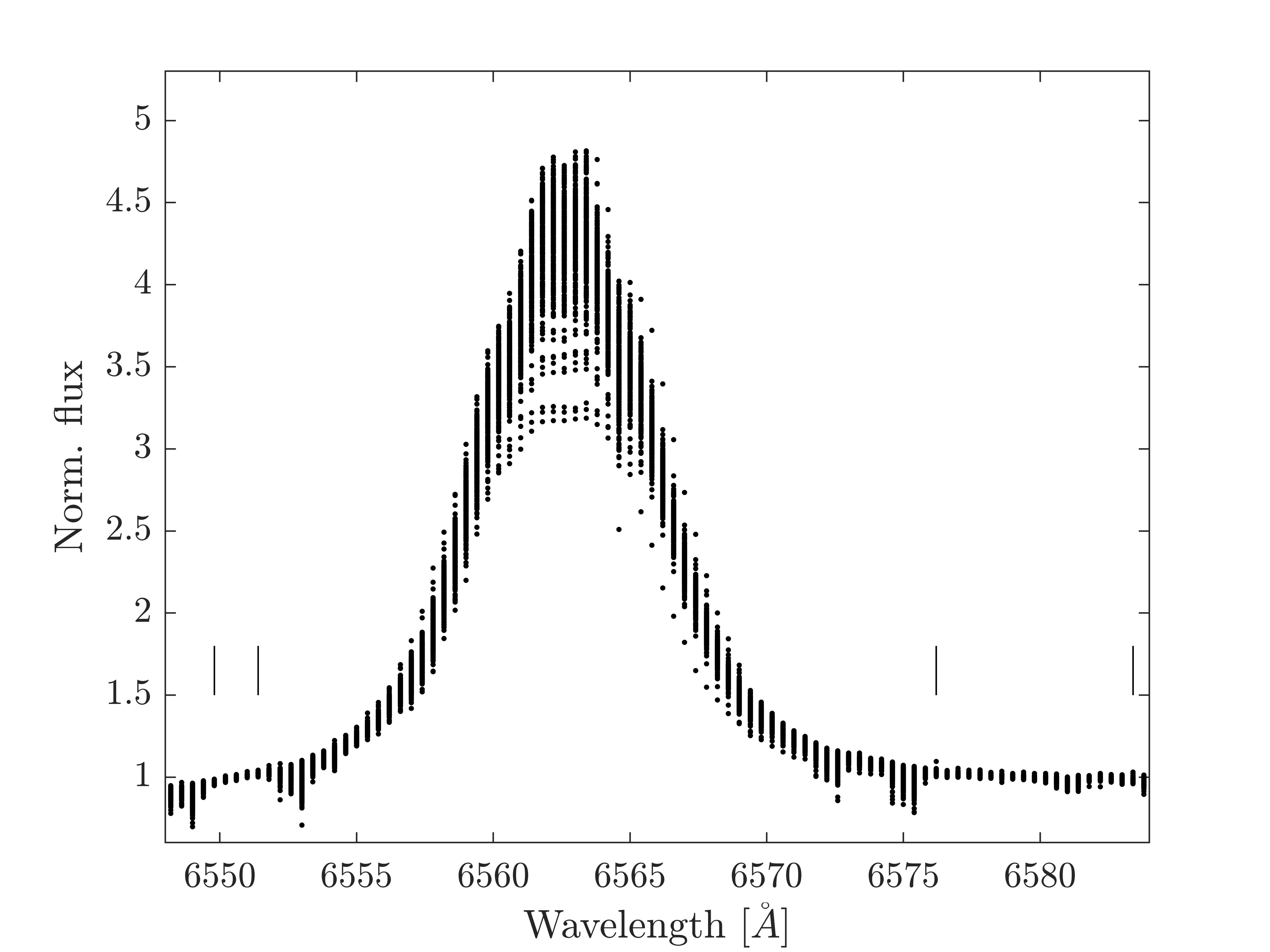}
			\caption{321 H$\alpha$ profiles from the BeSS Database are binned in wavelength with a bin size of 0.4\,$\AA$. All profiles are normalized and the thin vertical lines indicate areas used for normalization. The plot shows all profiles and they are therefore not resolved vertically. The H$\alpha$ line is fairly stable compared to many other Be stars and roughly has a flat-top shape.}
			\label{Fig:bindata}
		\end{figure}	

	\section{Observations and data reduction}
	\label{sec:obs}

	\subsection{Be Star Spectra Database (BeSS)}
\label{sec:BeSS_telescope}
 The BeSS Database\footnote{\url{ http://basebe.obspm.fr}} \citep[][]{Neiner2011} is ideal for studying spectral variations of Be stars on many timescales because it has a long time baseline of observations. The aim of this database is to collect spectra for all known Be stars and it is populated with data obtained by both professional and amateur observers. 

The spectra in this analysis were selected using the following criteria: 1) a minimal wavelength range from 6550 to 6575 $\AA$ and 2) a resolving power of 15,000 and above. The BeSS query system uses the resolving power in the FITS headers which is mostly automatically added by the respective data acquisition system. The observers can, however, also enter a number manually if something went wrong during the observations, for example, the calibration of the instrument (C. Neiner, personal communication, 2018). We used the BeSS entries of the resolving power but removed spectra that were obviously of lower resolution. A total of 389 spectra were downloaded on 22.11.17. 
After removal of spectra with saturated profiles or low resolving power, 371 spectra remained. 
The analysis was further limited to spectra observed after HJD 2453962.5 (15 Aug 2006) since there are only very few spectra taken earlier. This still means that the time span exceeds $\sim$11 years (2006-2017) and only 50 spectra were left out. Furthermore, all the spectra used by \cite{Harmanec2000} and most spectra used by \citet{Nemravova2012} have not (to our knowledge) been deposited in the BeSS Database. At most 13 of the 321 spectra in this work are also used by \citet{Nemravova2012}. This makes the analysis performed in this study not only different from previous investigations but it is also applied to an independent dataset.

The spectra were corrected to the heliocentric reference frame, binned in wavelength to 0.4 $\AA$, and normalized to the continuum. 
Caution was taken when normalizing since some of the spectra did not extend over a very large wavelength range and therefore did not include a long stretch of continuum. Also, regions affected by strong telluric lines were not used to define the continuum.
We selected two areas (one on each side of the H$\alpha$ peak) 
and used a linear fit of the two regions to normalize the spectra. The result and normalization regions can be seen in Fig.\,\ref{Fig:bindata}. 

Each wavelength bin contains relative fluxes as a function of time, i.e. a time series. The time series were further prepared, now in a time domain, to search for periodic variations. 
The data in each wavelength bin were divided by a fitted 2nd-degree polynomial to remove long-term (possibly non-periodic) trends. These trends can arise from several circumstances including the variety of instruments used to obtain the observations. Regardless, the de-trending does not remove any of the shorter variations i.e. shorter than 5 years. We then prewhitened the time series of every wavelength bin to reduce the yearly variations. This was done by calculating the amplitude spectrum as described in \cite{Kjeldsen_phd} and \cite{Frandsen1995} and then removing the desired frequency. 
We prewhitened for 1, 2 and 3\,yr$^{-1}$. There was a strong yearly variation in the data; however, removal of the 2 and 3\,yr$^{-1}$ contributions had a smaller effect and did not change the outcome significantly.

		\begin{table}
		\centering
		\caption{Overview of the SMEI observations}
		\label{tab:SMEI:datasets}
		\begin{tabular}{c l c c}
			\hline 
			Camera & JD start-end & No. of data    & No. of data   \\ 
			 
			& $-$2,450,000 &  points before  & points after  \\
			&              &  reduction        & reduction \\
			\hline 
			1	& 2867.4-5833.3 & 10055 & 9787 \\ 
			2   & 2673.5-5810.1 & 29891 & 28855 \\ 
			3   & 2713.1-5681.2 & 4288 & 0 \\ 
			\hline 

		\end{tabular} 
		
	\end{table}

	\subsection{Solar Mass Ejection Imager (SMEI)}
	\label{sec:SMEI_telescope}

	\begin{figure}
		\centering
		\includegraphics[width=8.8cm]{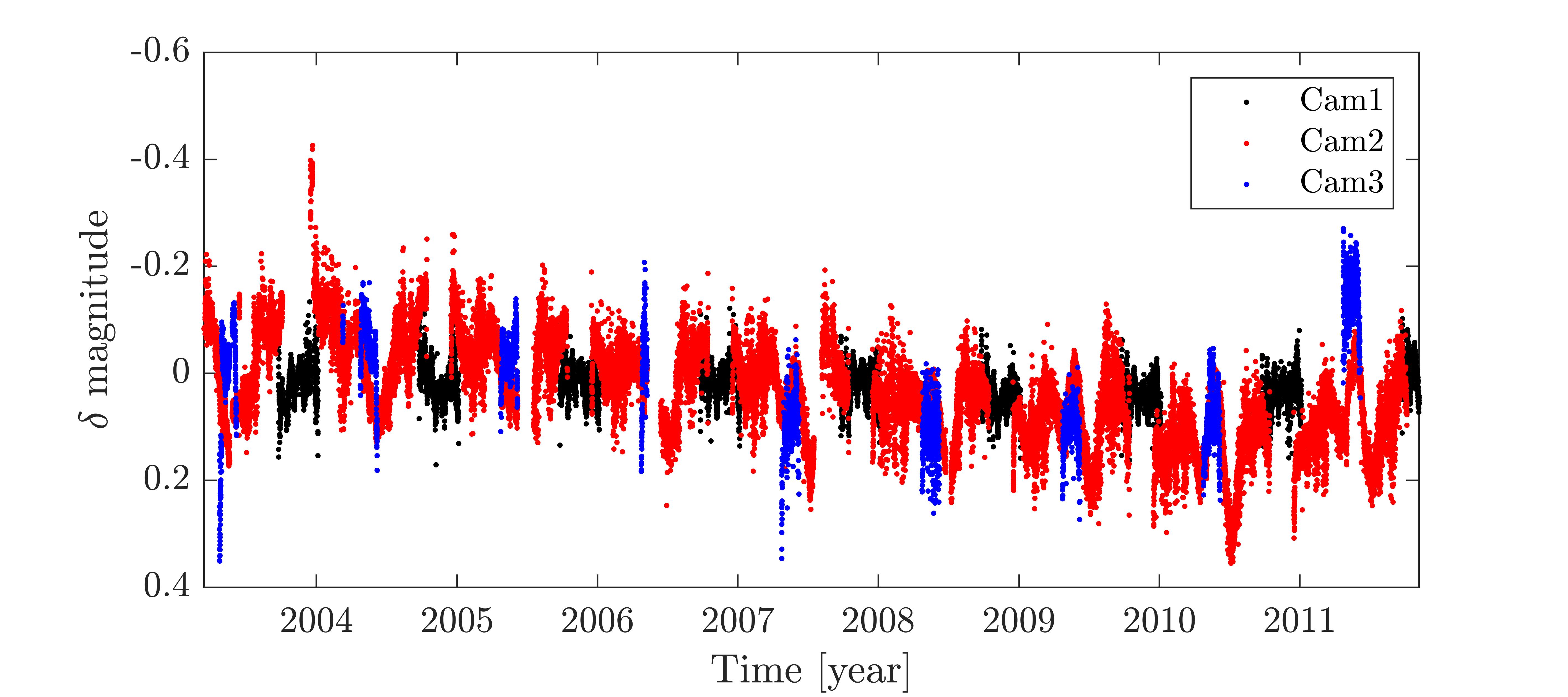}
		\caption{Reduced data from SMEI Cameras\,1 and 2.  The data from Camera\,3 are not reduced and only shown for comparison; they are not used in the time-series analysis.}
		\label{Fig:time_SMEI}
	\end{figure}

	SMEI was launched onboard the Coriolis satellite in 2003 to monitor space weather \citep{Jackson2004}. Its purpose was to detect Thomson-scattered sunlight from within the heliosphere volume. It was designed to have a lifetime of 5 years but was operated until September 2012. It had 3 cameras, each with a field of view of 3$\times$60 deg$^2$, and was placed in a near-Earth orbit with a period of $\sim$100 minutes. The combination of the three cameras gave a 160$^\circ$ arc sweep of the sky. The aperture of each camera was approximately $1\times2$ cm$^2$.  
	
	To reach the required photometric precision when mapping the Thomson scattered light across the sky, all stars brighter than 10th magnitude had to be individually fitted and removed. The point spread function (PSF) used for the extraction had a full width of about 1 deg and varied in width depending on the object's path through the field of view. This is a large diameter and attention should be paid when ascribing flux variations to a star because multiple bright stars can be located within the same PSF \citep{Hick2007}. SIMBAD \citep{Wenger2000} lists one source brighter than a threshold of 7th magnitude within 0.6$^\circ$ of \gC. This is the triple system \object{HD 5408} which has a magnitude of $\sim$5.5 while \gC has a magnitude of $\sim$2.5. Orbital periods for HD 5408 are known to be around 83 and 5 years \citep{Tokovinin1997}. 
	
	In parallel with the SMEI observations, \gC was observed with the APT which has a photometric aperture of at most 2 arcmins. 
	Within this radius, there are no stars brighter than 13th magnitude which is not expected to influence the results. 
	Since the 0.82\,d$^{-1}$ frequency, found in this work with SMEI, has already been detected with APT \citep{Smith2006,Henry2012} (and HD 5408 is not included in the APT aperture) the 0.82\,d$^{-1}$ frequency is not due to HD 5408. 
	
	The UCSD pipeline-extracted \citep{Hick2005}  stellar fluxes \citep[in SMEI magnitudes;][]{Buffington2007} are available on the web\footnote{\url{http://smei.ucsd.edu/new_smei/data&images/stars/timeseries.html}}. 	
	 Unfortunately, the web site only provides a single combined dataset for each star with no possibility of distinguishing between the different cameras. The data used in this work were kindly provided by B. Jackson (2018, private communication to DB) and contain information about the cameras, making it possible to separate the datasets. This turned out to be important as will be discussed below and in Sect.\,\ref{sec:SMEI_analysis}.
	
	\gC was observed from 2004 to 2011 with all three cameras. The data were reduced using the pipeline but had to be further processed due to the large noise and large seasonal variations (see Fig.\,\ref{Fig:time_SMEI}). This was done using a moving median assuming a normal distribution and a threshold of 4$\sigma$. The observation periods and number of data points can be seen in Table\,\ref{tab:SMEI:datasets}. Camera\,3 was the most sunward camera and damaged during the mission due to radiation from the sun. While Cameras 1 and 2 were kept at a temperature around $-$30 $^\circ$C to reduce noise, Camera\,3 operated at temperatures between $-$2 and $-$15 $^\circ$C which lowered the sensitivity \citep{Webb2006}.
	For this reason and because the number of data points from Camera\,3 is relatively low, all the data from this camera have been discarded. 
	
	Lastly, the data were prewhitened for 1, 2, 3, 4, 5 and 6\,yr$^{-1}$ and 1, 2, 3, and 4\,d$^{-1}$.  
	The reduced, prewhitened and normalised time series is shown in Fig.\,\ref{Fig:time_SMEI}. Data from Camera\,3 are presented only for comparison and will not be used in further calculations.

	\begin{figure}
	\centering
	\includegraphics[width=8.8cm]{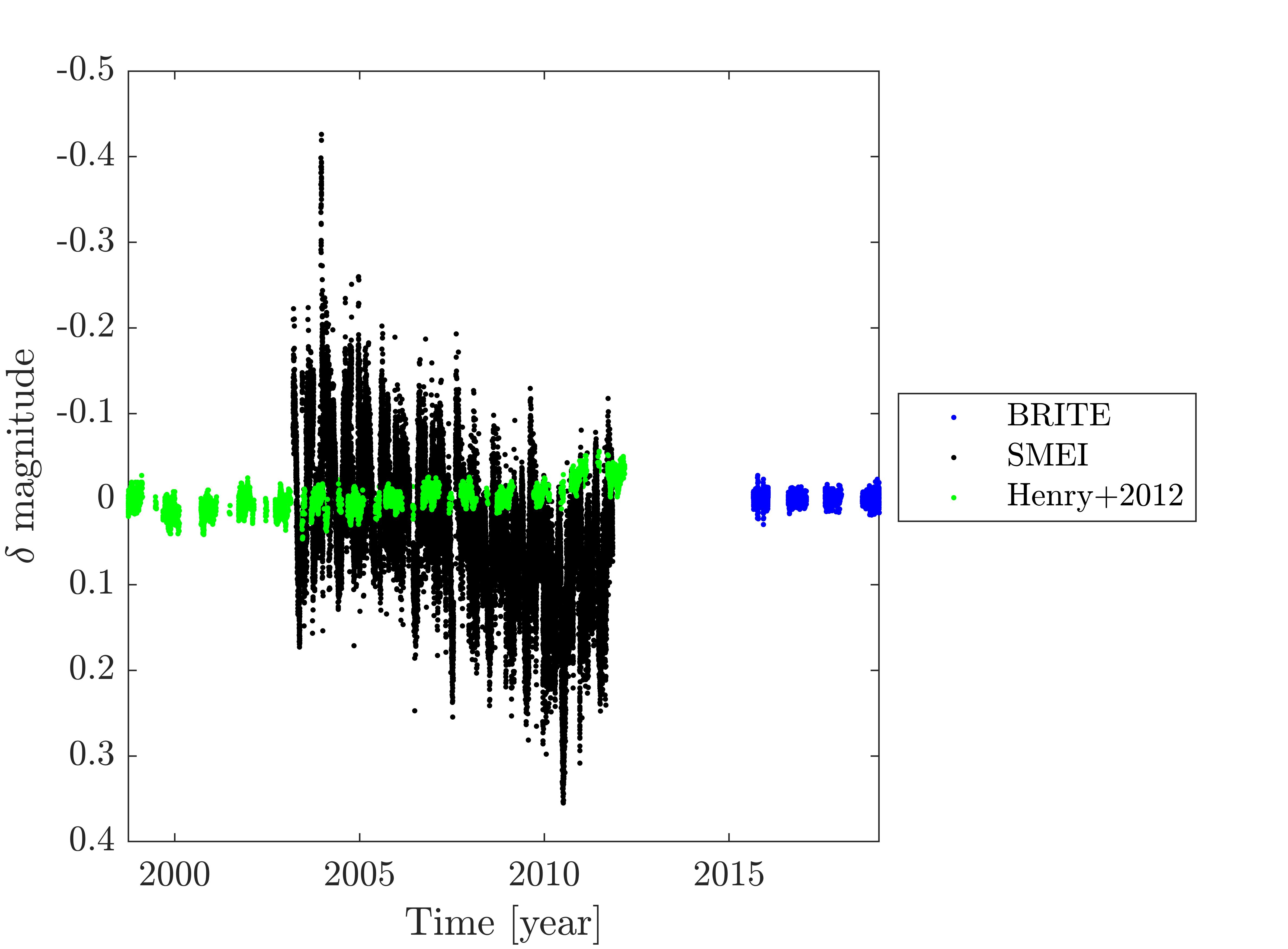}
	\caption{Time series of BRITE (blue), SMEI (black) and APT (green) photometry. The mean magnitude of SMEI and BRITE is 0 by construct. The mean magnitude of the APT data has been shifted to zero compared to the data provided on the web. The APT data were analysed in detail by \cite{Henry2012}. The figure shows that the noise in the SMEI satellite data is significantly higher than in the APT and BRITE measurements. }
	\label{Fig:All_timeseries}
	\end{figure}

	\begin{table*}
	\caption{Overview of the BRITE observations}
	\label{tab:BRITE:datasets}
	\centering
	\begin{tabular}{l c c c l c c c}
		\hline   
		Satellite name & Orbital period & Year & Contig. time  & JD start-end & Range in  & No. of  & No. of TSA  \\ 
		(acronym)&  [min] &  & [min] & -2,457,000 &  CCDT [$^\circ$C] & exposures & data points \\
		\hline 
		BRITE-Austria  & 100.4 & 2015  &  4.7-19.3   & 263.5-321.6 & 17.7-30.9 & 17349 & 501 \\ 
		(BAb)		   &   & 2016 & 5.8-14.6 & 608.3-751.5 & 16.2-29.1 & 24837 & 752 \\ 
		&   & 2017 & 3.0-14.8 & 973.4-1,133.6 & 17.9-33.8 & 13086 & 470 \\ 
		&   & 2018 &  2.6-9.8  & 1339.9-1435.8  &  18.0-31.8   &  4569   &   121  \\   
		
		BRITE-Lem	   & 99.6 & 2015 & 7.5-20.3 & 294.9-312.2 & 32.1-39.4  & 9910 & 163 \\ 
		\multicolumn{8}{l}{(BLb)} \\	
		
		BRITE-Heweliusz & 97.1 & 2015 & 4.1-19.6  & 266.4-313.3 & 8.5-20.8 & 5226 & 143 \\ 
		\multicolumn{8}{l}{(BHr)} \\

		BRITE-Toronto	& 98.2 & 2015 & 4.4-16.6  & 360.8-408.5 & 12.7-23.1 & 17111 & 632 \\ 
		(BTr)		   &   & 2018 &  5.4-10.9 & 1410.0-1505.5 & 16.0-26.5 &  32460  & 1027 \\

		UniBRITE	& 100.4 & 2016 & 4.7-14.9 & 645.1-786.1 & 17.6-31.5 & 29934 & 838 \\  
		\multicolumn{8}{l}{(UBr)} \\	
		
		\hline	\multicolumn{8}{l}{\textbf{Note:} Suffixes 'r' and 'b' refer to passbands red and blue, respectively. 'Contig. time' denotes the typical contiguous time} \\	
		\multicolumn{8}{l}{interval per orbit during which exposures were made. CCDT is the temperature of the detector. TSA (time series analysis) }\\
		\multicolumn{8}{l}{data points is the number of data points formed by an orbital average of the exposures and after data reduction. }\\
	\end{tabular} 
	
    \end{table*}

	\begin{figure*}
	\centering
	\includegraphics[width=18cm]{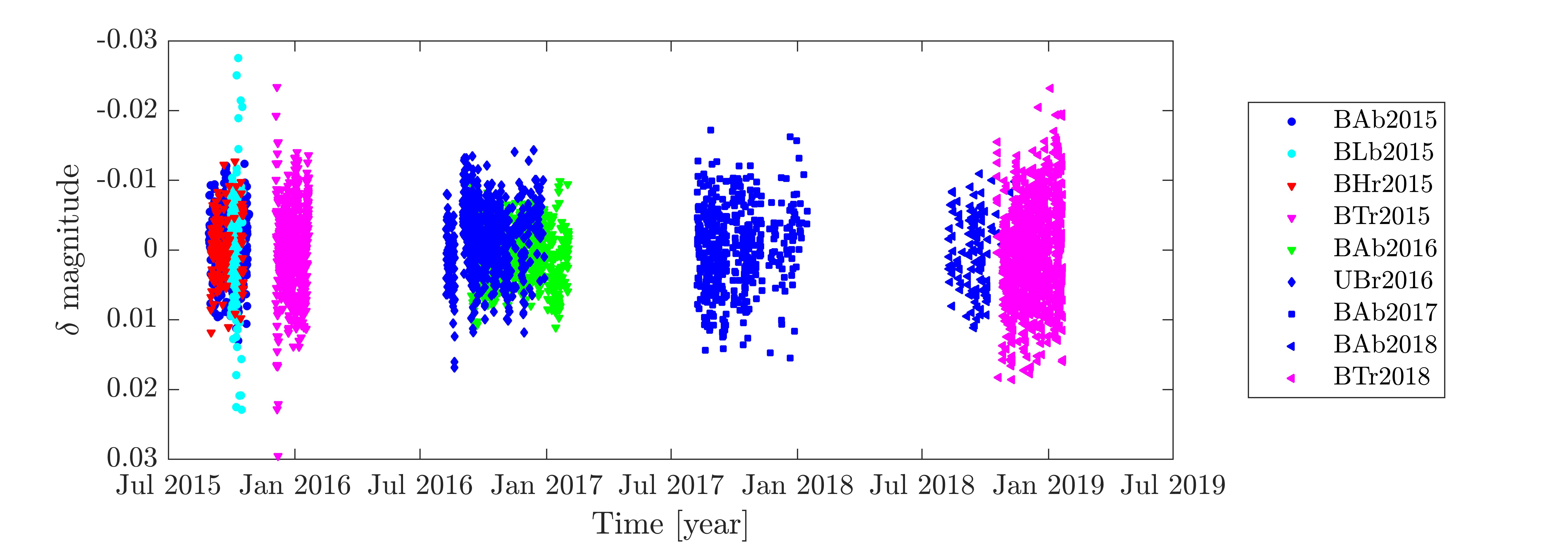}
	\caption{Time series of BRITE photometry. The mean magnitude of the individual datasets is 0 by definition. }
	\label{Fig:BRITE_timeseries}
    \end{figure*}

			\subsection{BRIght Target Explorer (BRITE)}
	\label{sec:BRITE:telescope}	
	BRITE is a constellation of five nano-satellites operated by three countries, namely Austria, Canada, and Poland \citep{Weiss2014}. The first satellite to be launched in 2013 was named UniBRITE (UBr) followed by BRITE-Austria (BAb), BRITE-Lem (BLb), BRITE-Toronto (BTr) and BRITE-Heweliusz (BHr). A report on pre-launch and in-orbit tests can be found in \cite{Pablo2016}. 
	The "r" ("b") in these acronyms stands for a red 550-700 nm (blue 390-460 nm) filter. The satellites have a field of view of 20$\times$24 deg$^2$ and a 3 cm aperture and similar to SMEI they were placed in a near-earth orbit with a period of about 100 minutes. The PSF varies across the field but has a size of about 3-6 arcmin$^2$ \citep{Popowicz2017}. At a radius of 6 arcmins from \gC  there are no stars brighter than $\sim$11th mag. 
	The goal of these satellites is to acquire simultaneous and nearly continuous dual-broadband photometry of bright stars for up to half a year. 
	In contrast to the SMEI project, the BRITE satellites are specifically designed to observe bright stars and the metadata are much more detailed. 
	
	The BRITE data have been extracted using aperture photometry and were pipeline reduced prior to the reduction described below. A description of the pipeline reduction can be found in \cite{Popowicz2017} and is the same for all the observation runs presented here. Due to the slightly different set-ups (mostly resulting from field re-acquisitions)  between the satellites, we further decorrelate and reduced each dataset individually. The reduction can also depend on the specific star requiring this individual computation.
	Furthermore, during an observing run, a single satellite can have different set-ups. These are then reduced separately and combined at the very end. The needs for decorrelations and typical methods are further described in the official "Cookbook" \citep{Pigluski2018}. As suggested in the Cookbook, we use orbital averages of the data points. This means that the Nyquist frequency is $\sim$7.2\,d$^{-1}$ which is sufficient for our analysis. For each setup, the mean magnitude is set to zero because the BRITE satellites only deliver relative photometry.
	After the reduction, the data were prewhitened for 1, 2, 3, 4, 5 and 6\,yr$^{-1}$ and 1, 2, 3, and 4\,d$^{-1}$. This was done although no significant peaks at these frequencies were seen in the amplitude spectra.

	In Table\,\ref{tab:BRITE:datasets} information about the BRITE data is given. The reduced and prewhitened BRITE time series in context of SMEI and APT \citep{Henry2012} is shown in Fig.\,\ref{Fig:All_timeseries}. The BRITE time series alone is presented in Fig.\,\ref{Fig:BRITE_timeseries}.

	\section{Results of the analysis}
		Similar to Sect.\,\ref{sec:obs} this section is organized by observing facility. Some discussions of the frequencies are therefore repeated. 
	\label{sec:analysis}
	\subsection{Frequencies in BeSS spectra}
	\label{sec:BeSS}
\begin{figure}
	\centering
	\includegraphics[width=8.8cm]{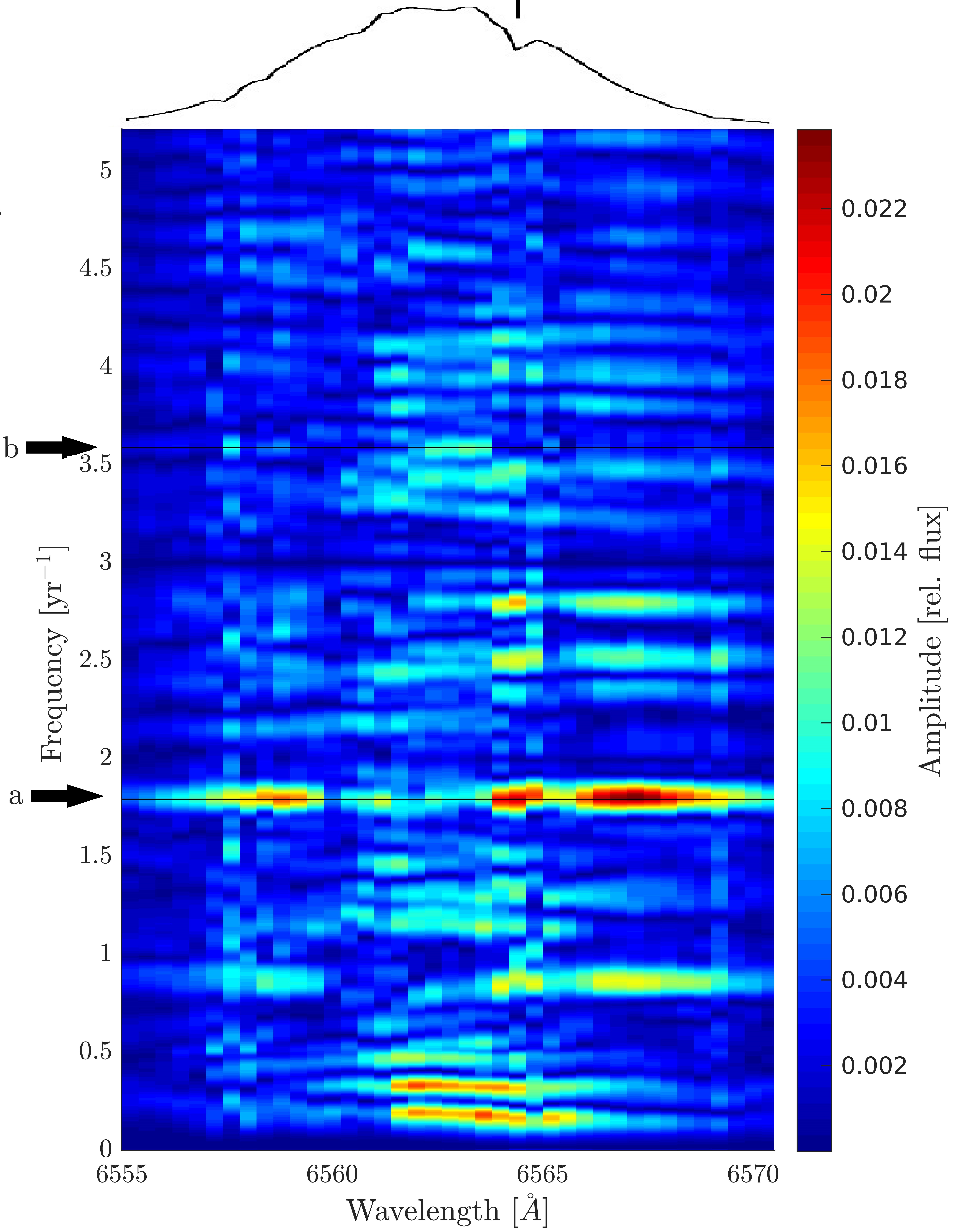} 
	\caption{Wavelength-resolved amplitude spectrum for \gC across the H$\alpha$ profile. An example of the H$\alpha$ profile is shown on top as reference. Arrows mark regions of interest: \textit{a}) stellar orbital frequency, \textit{b}) twice the stellar orbital frequency. The signature at $\sim$6564.5\,\AA\xspace (marked on the profile) is caused by telluric contamination.}
	\label{fig:GamCas_Surface2}
	\end{figure}

	The amplitude spectrum of the de-trended and prewhitened BeSS data was calculated for all wavelength bins individually and is presented in Fig.\,\ref{fig:GamCas_Surface2}. 
	On top of the amplitude spectrum, an example of an H$\alpha$ line is shown for easier cross identification between line structure and wavelength. The dash at $\sim$6565 $\AA$ indicates a telluric line and this region should be interpreted with caution. Arrows point at frequencies of interest which are discussed below. 
	
	The largest amplitude is marked with arrow \textit{a}. The wavelength bins between 6556 and 6560 as well as between 6565 and 6569 $\AA$ were fitted with a sinusoid  resulting in a mean period of 202.9(5) days (uncertainty from standard deviation). This agrees with the known binary orbital period of 203.52(8) days that has been reported by others \citep[see for example][]{Harmanec2000,Nemravova2012}. These papers measure the periods using the radial velocity of the emission line wings.  
	As this method has already been used extensively on this star, we do not repeat it in this work.  
	By examining the flux change in each wavelength bin (as is done here) instead of the radial velocity, the structural change of the disk can be probed. This is due to the strong correlation between the structure of the H$\alpha$ emission and that of the disk \citep{Horne1986}. The frequency is dominant in the wings because a shift in a flat top does not lead to a significant signal. 
	
	The blue/red asymmetry of the amplitude in the wings is, on the other hand, a remarkable result. 
	\cite{Panoglou2016} show simulations that demonstrate how the companion can induce or at least influence a spiral structure in the disk so that the rotation of the companion and the spiral becomes phase-locked. This asymmetry could, therefore, arise because of the observational difference between the concave and convex part of the spiral arm. For example, if the optical thickness of the gas is different in the denser front part of the spiral arm compared to the gas trailing behind it, the reprocessed light from the star is also different. Given a one-armed structure, one would equally often observe the back and front side of the spiral arm but at different luminosities. This would give an equal frequency of the two sides of the arm but with different amplitude, as is seen in Fig.\,\ref{fig:GamCas_Surface2}. According to simulations, a two-armed structure could form \citep{Panoglou2016} which could produce a signal at double the orbital frequency (arrow \textit{b}, Fig.\,\ref{fig:GamCas_Surface2}). We do not observe such a signal and we discuss why this might be the case in  Sect.\,\ref{sec:discussion_spiral}.

	To check the robustness of the orbital signal, we split the data sample into two halves - by time as well as by randomly selecting the data. In all four cases, the orbital frequency is still visible although (naturally) not as prominent.
	
\begin{figure}
	\centering
	\includegraphics[width=8.8cm]{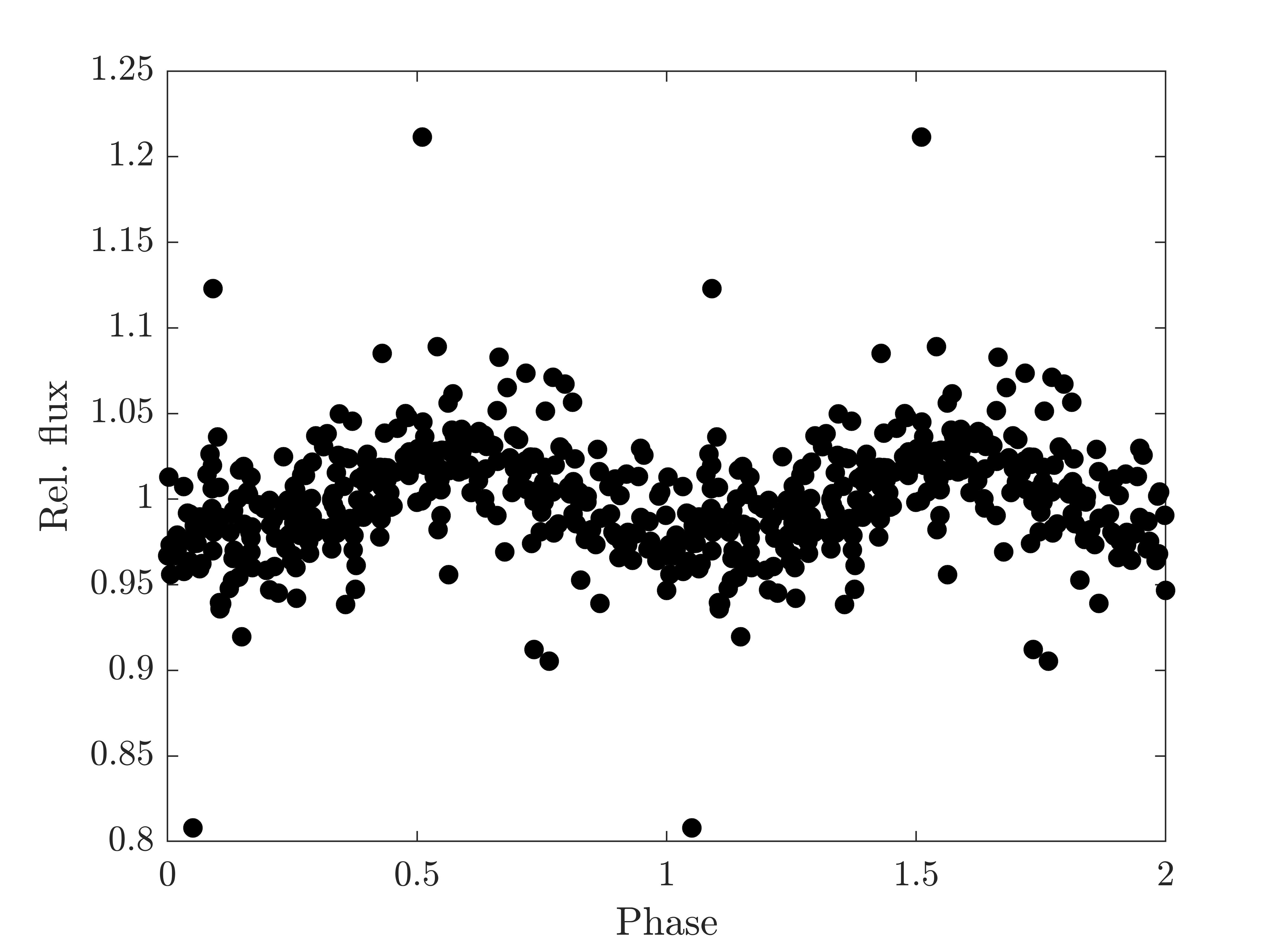}
	\caption{Phase folded time series of H$\alpha$ fluxes at wavelength bin 6567.1 \AA. The data are folded with the orbital period and are repeated for clarity.}
	\label{fig:GamCas_fold2}
	\end{figure}

	The time series was folded with the orbital period and Fig.\,\ref{fig:GamCas_fold2} shows that it is well defined and that the data are evenly distributed over the phases.

	\subsection{Frequencies in SMEI photometry}
	\label{sec:SMEI_analysis}
\begin{figure}
		\centering
		\includegraphics[width=8.8cm]{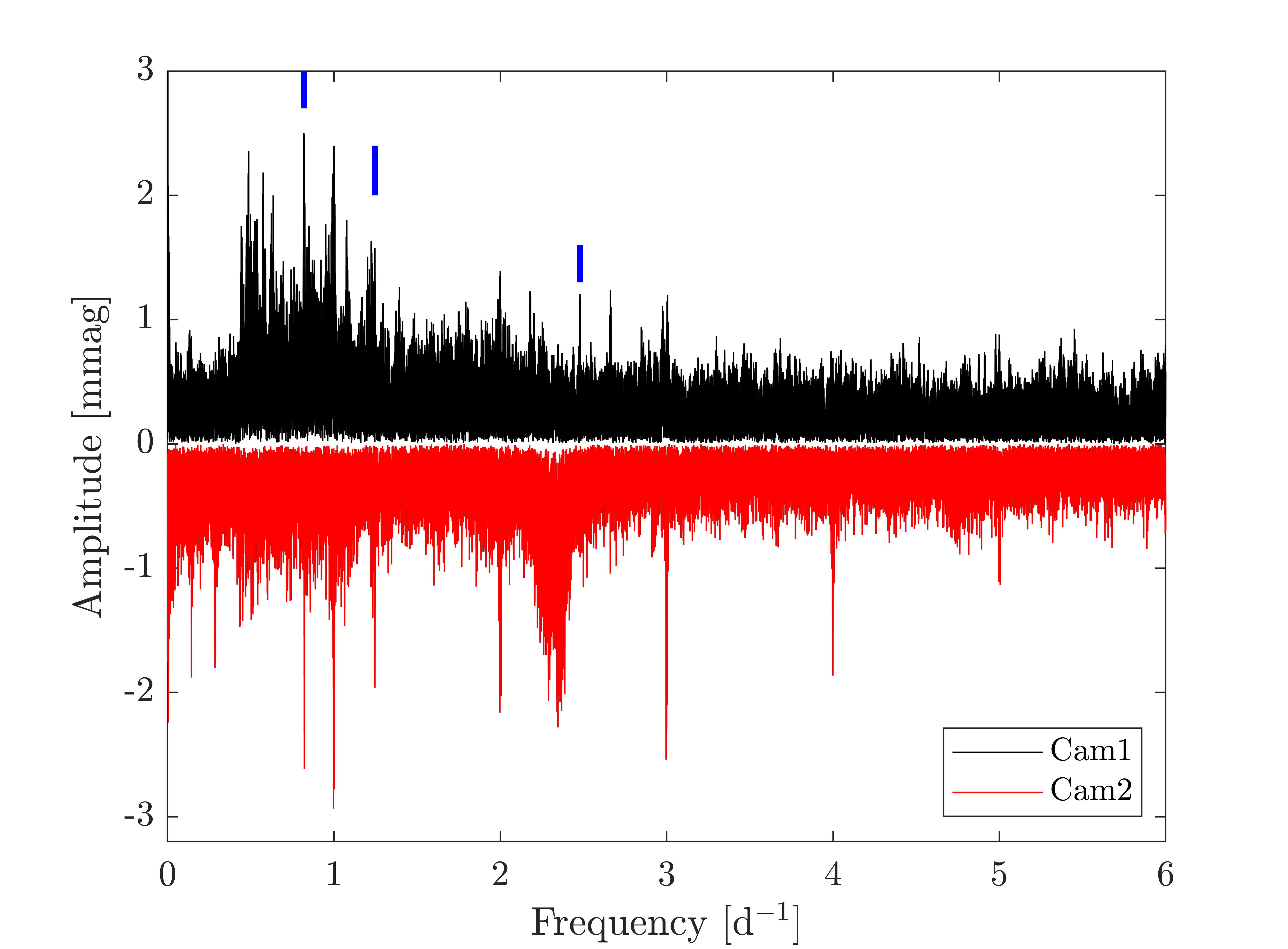}
		\caption{Amplitude spectra of the two SMEI time series. That from Camera\,2 has been flipped to negative values for clarity. Frequencies 0.82, 1.25 and 2.48\,d$^{-1}$ are marked.}
		\label{fig:SMEI_Power12}
	\end{figure}

	The SMEI data display slow trends (clearly seen in Fig.\,\ref{Fig:All_timeseries}) that can affect the amplitude spectrum and we have therefore applied a high-pass filter to reduce their effects on the noise at higher frequencies. 
	The amplitude spectra for Camera\,1 and Camera\,2 of the SMEI data can be seen in Fig.\,\ref{fig:SMEI_Power12}.  
	
	It is clear that some of the major features correspond to 1, 2 and 3\,d$^{-1}$ which have (most likely) nothing to do with the star but rather (the local environment of) the satellite or from light scattered off clouds and ice on Earth. These features have been reduced through the prewhitening process but they are not entirely gone. This could be because the prewhitening is only done on one peak in the amplitude spectrum and its associated features in the window function. This means that even though the code takes the structure of the data into account, the signal is not perfectly sinusoidal and it might not be removed completely. It is, however, not important to fully remove all these features as long as one is careful not to attribute them to the star. We tested a more detailed prewhitening (with more peaks removed) but it did not affect the detections of the stellar frequencies in any significant way. 

	In Camera\,2 data there is a large bump at 2.3-2.4\,d$^{-1}$. Such a bump is seen in many SMEI datasets (although not always at the same frequencies) and is, therefore, considered instrumental (Pigulski, private communication).

\begin{figure}
		\centering
		\includegraphics[width=8.8cm]{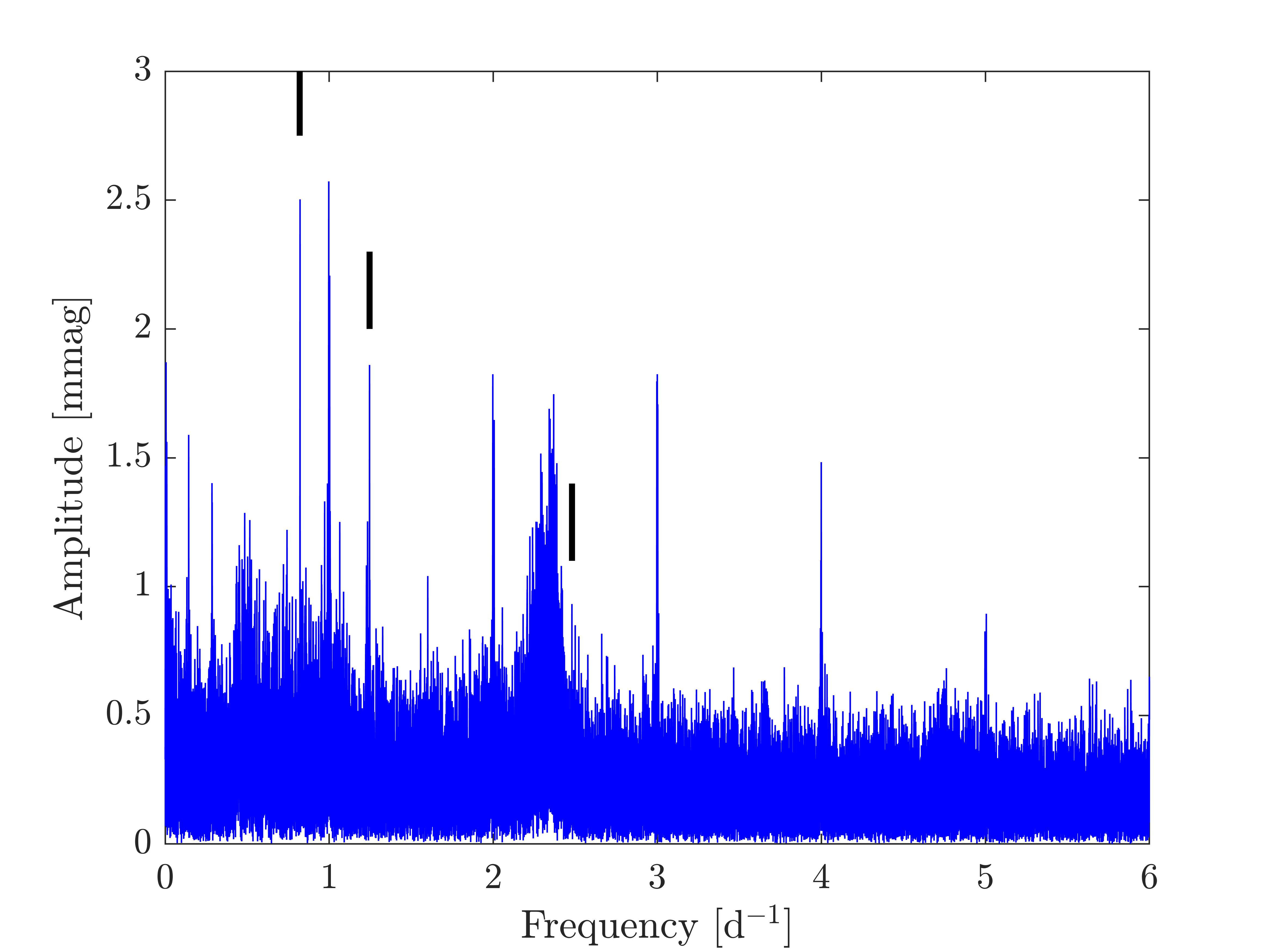}
		\caption{Amplitude spectrum of the combined SMEI data (Camera\,1 + Camera\,2). Frequencies 0.82, 1.25 and 2.48\,d$^{-1}$ are marked.}
		\label{fig:SMEI_Power}
	\end{figure}

    The most significant non-instrumental signal in both spectra is at 0.82\,d$^{-1}$. By combining the time series of the two cameras we obtain a time series with smaller gaps and a tighter constraint on the frequency. The amplitude spectrum from SMEI Cameras 1 and 2 combined is shown in Fig.\,\ref{fig:SMEI_Power}. For frequency determination throughout the rest of this work, we use Period04 \citep{period04} which is the accepted standard in the field of stellar pulsation analysis. Period04 calculates uncertainties as described by \citet{Breger1999} and \citet{Montgomery1999}. Analysing the two datasets separately we obtain a frequency of 0.82200(2)\,d$^{-1}$ for Camera 1 and 0.82217(2)\,d$^{-1}$ for Camera 2. These frequencies agree within 4.3$\sigma$ which we accept as consistent because the uncertainties returned by Period04 assume normally distributed noise which is probably not true for SMEI data. To achieve a higher resolution and a more precise frequency value, the two datasets were combined into one in which we find the frequency to be 0.82215(1)\,d$^{-1}$.
    This is not the first time that this frequency has been reported. \cite{Smith2006} and later \cite{Henry2012} observed this variation using a 15-year-long time series from the ground-based APT. They found a frequency of 0.82250(2)\,d$^{-1}$ which was recently revised by \cite{Smith2019} to 0.82247(2)\,d$^{-1}$. The difference between our and their result is approximately 1.5 cycle in 12 years, which corresponds to the length of the data string, and this could explain the discrepancy. 
	
	Unfortunately, the large scatter in the data and strong aliases mean that the comparatively small amplitude of the 0.82\,d$^{-1}$ frequency is not visible in the light curve - even when the data are folded with this frequency or binned in phase. 
	
\begin{figure}
	\centering
	\includegraphics[width=8.8cm]{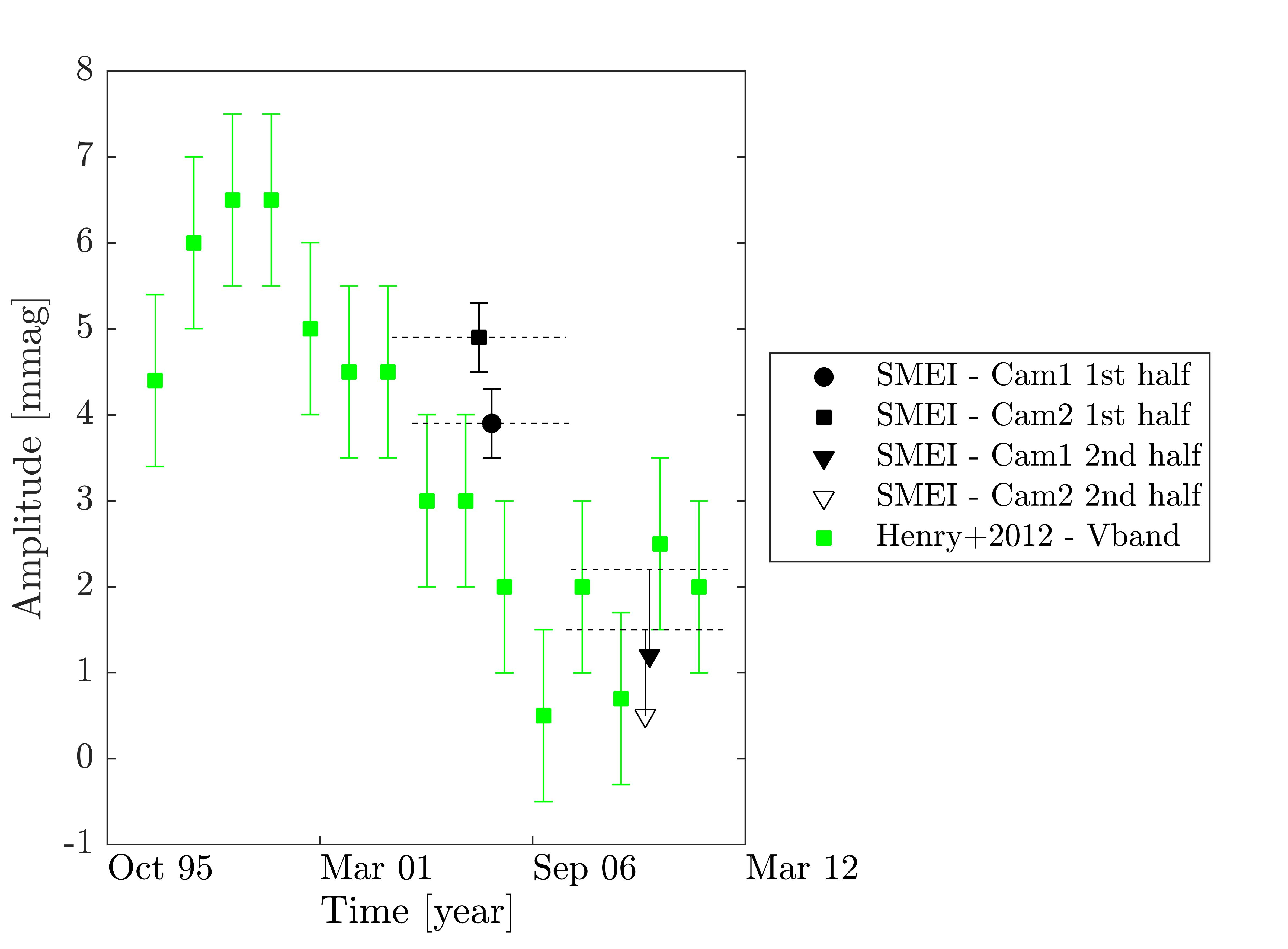}
	\caption{Comparison of amplitudes of the 0.82\,d$^{-1}$ frequency determined in this work (SMEI) and reported by \cite{Henry2012} (APT). The two SMEI datasets were split into two subsets. The first point for each camera is a measurement of the amplitude of the signal while the second point (triangle) is an upper limit. The dashed lines indicate the time span of the two SMEI subsets. When BRITE observed the star in 2015-2018 this frequency was not recovered with a detection limit of $\sim$1 mmag.}
	\label{Fig:SMEI_amplitude}
    \end{figure}

	\cite{Henry2012} show that the amplitude varies with time. To test this finding, we split the time series into two halves for each camera. The amplitudes of the signal in the four subsets are shown in Fig.\,\ref{Fig:SMEI_amplitude} along with the results from \cite{Henry2012} for comparison. We read off the values from their Fig.\,7 instead of redoing their analysis where they preformed highly sophisticated  multi-level corrections for different time-dependent slow variations. For the first half of the time series, we could confidently retrieve the frequency. In the second half of the time series, the noise level of the SMEI data was too large and we have only been able to provide upper limits of the amplitude. These are calculated as 3 times the mean strength of all peaks in 0.2-c/d frequency windows around 0.82 c/d. 
	Although this only provides 4 data points, the combination with the \cite{Henry2012} data shows very well the changing amplitude of this variation. 
	
	In the following section, we show that a new frequency of 2.48\,d$^{-1}$ is present in the BRITE data. With this knowledge, we have searched for this in the SMEI data as well. From Fig.\,\ref{fig:SMEI_Power12} it is possible that such a frequency is present at least in Camera\,1 data. Camera\,2 data do not show this frequency at a significant level. The 2.48 c/d feature lies outside the broad (probably instrumental) bump between 2.3 and 2.4 c/d.  Therefore, this contamination should not be the explanation of the absence of the 2.48 c/d frequency from the Camera\,2 data.  However, we do not have another explanation. In the combined data (Fig.\,\ref{fig:SMEI_Power}) only a weak signal appears at the expected frequency. We find the frequency to be 2.47951(4)\,d$^{-1}$. There is no sign of this frequency in the window function but we cannot completely disregard that it is an annual alias of the aforementioned 0.82\,d$^{-1}$ since SMEI data, in general, have strong annual trends. Nevertheless, we regard it as genuine because it is present in BRITE data.  The 2.48\,d$^{-1}$ frequency was first reported by \cite{Baade_conf}. It was also reported for the contemporaneous APT data by \cite{Smith2019}, who, however, does not state in  which time interval nor at what amplitude this is seen. 
	
    Although data from both cameras appear to have other signals as well, the only signal that possibly exists in both datasets is at 1.24583(2)\,d$^{-1}$. While clearly present in Camera\,2 data, it is not certain if it is in Camera\,1 data. In the combined data it appears much more significant. The 1.25\,d$^{-1}$ signal was first found by multiplying the amplitude spectra of the two cameras which emphasizes signals and eliminates non-common noise spikes. We note that this is a different method than the combination of data and it was not used for analysis because it is not a reliable detection method. It only gives a first suggestion of what the two amplitude spectra have in common and any real signal should show up here. However, the latter is only a necessary and not a sufficient condition. With this method, the 0.82 \,d$^{-1}$, 1.25\,d$^{-1}$ and 2.48\,d$^{-1}$ frequencies were clearly present while all other possible contenders disappeared. 
	
	Lastly, we searched for the orbital period of the companion in an attempt to detect photometric changes in the lightcurve due to the possible spiral structure. Due to the large noise levels at low frequencies, we did not find any significant signal at this frequency.

\subsection{Frequencies in BRITE photometry}
\label{sec:BRITE:analysis}
\begin{figure*}
	\centering
	\begin{subfigure}[b]{0.45\textwidth}
		\includegraphics[width=\textwidth]{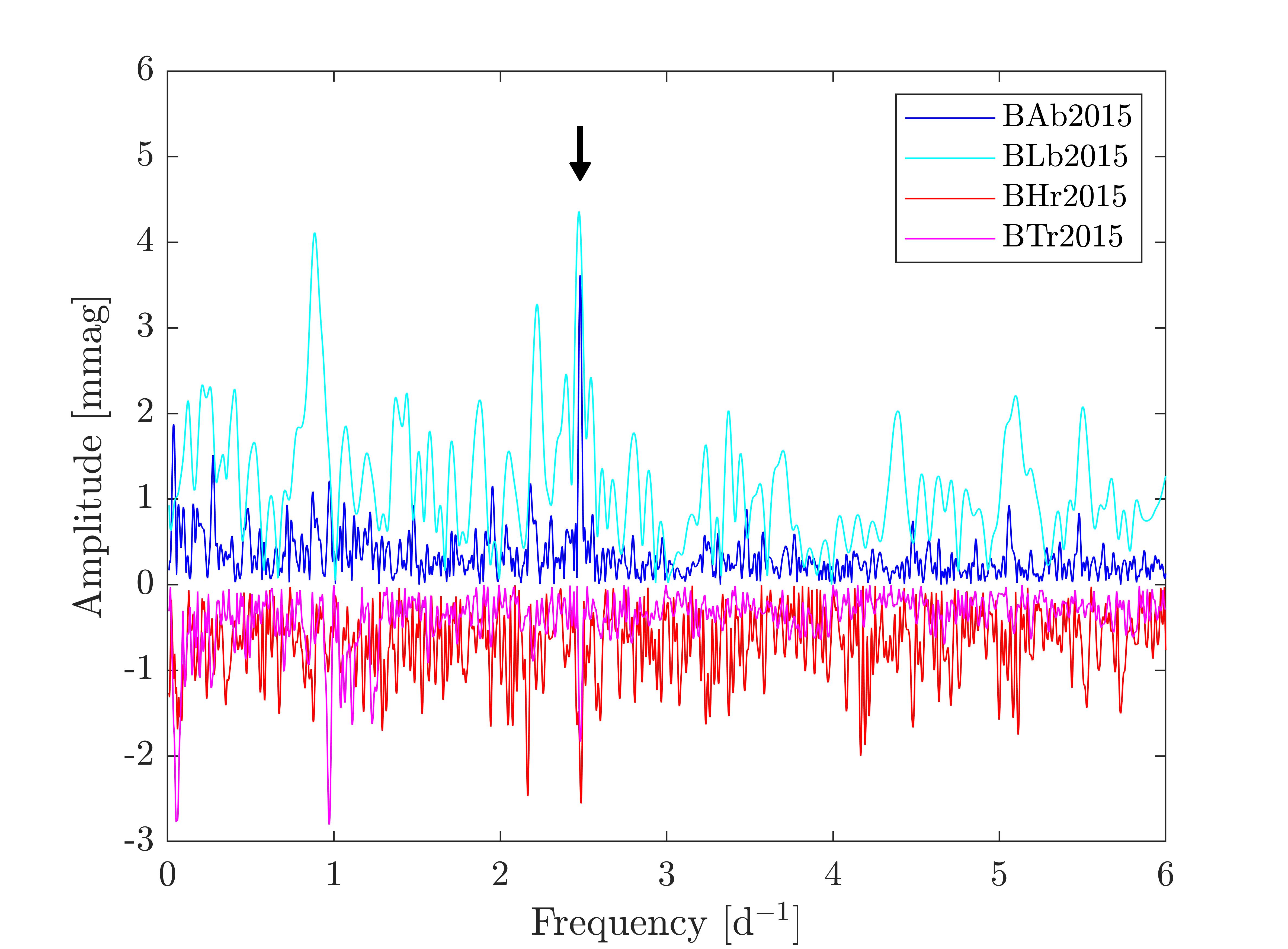}
		\caption{}
		\label{Fig:Power2015}
	\end{subfigure}
	~ 
	\begin{subfigure}[b]{0.45\textwidth}
		\includegraphics[width=\textwidth]{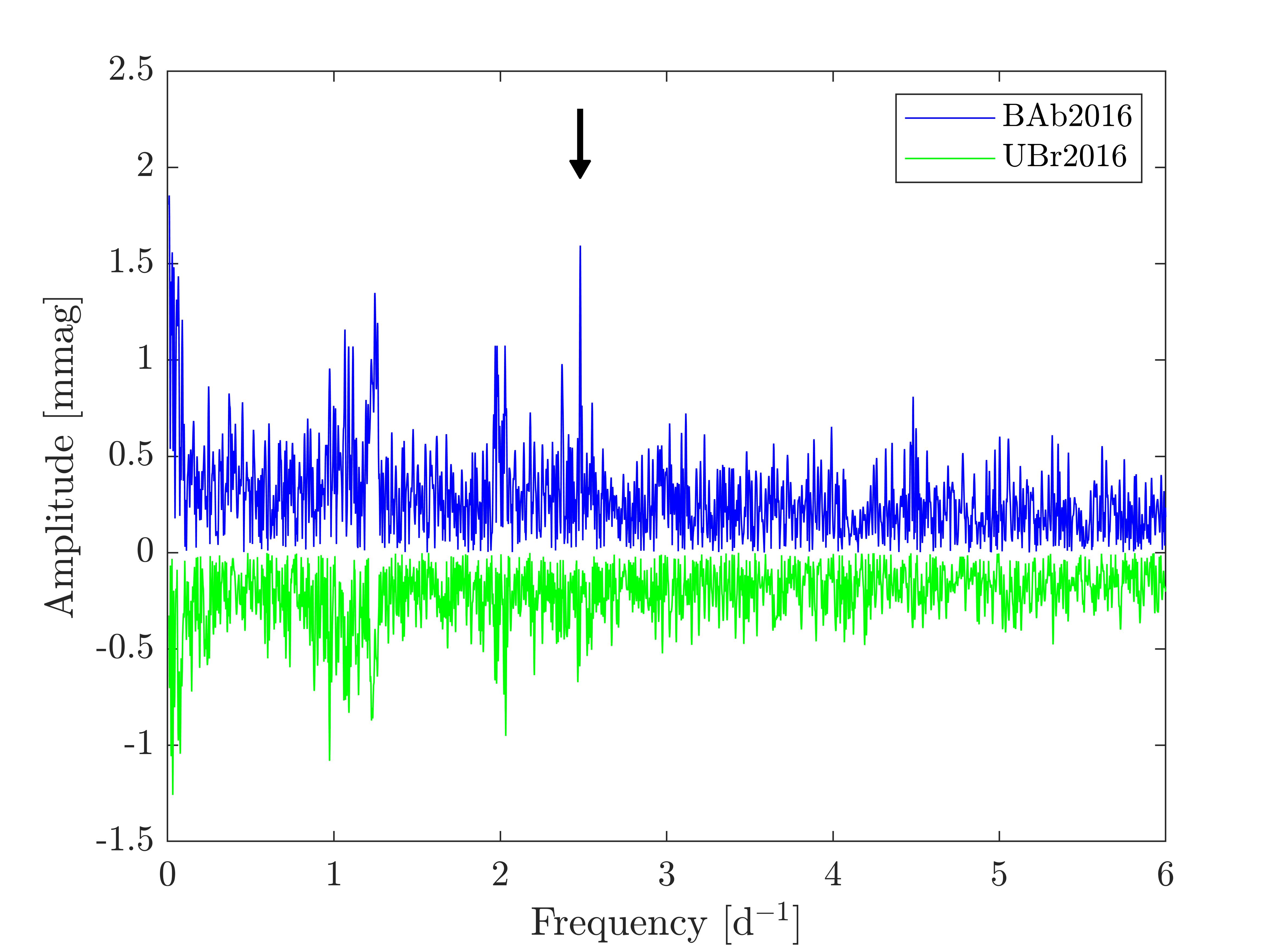}
		\caption{}
		\label{Fig:Power2016}
	\end{subfigure}
	
	\begin{subfigure}[b]{0.45\textwidth}
		\includegraphics[width=\textwidth]{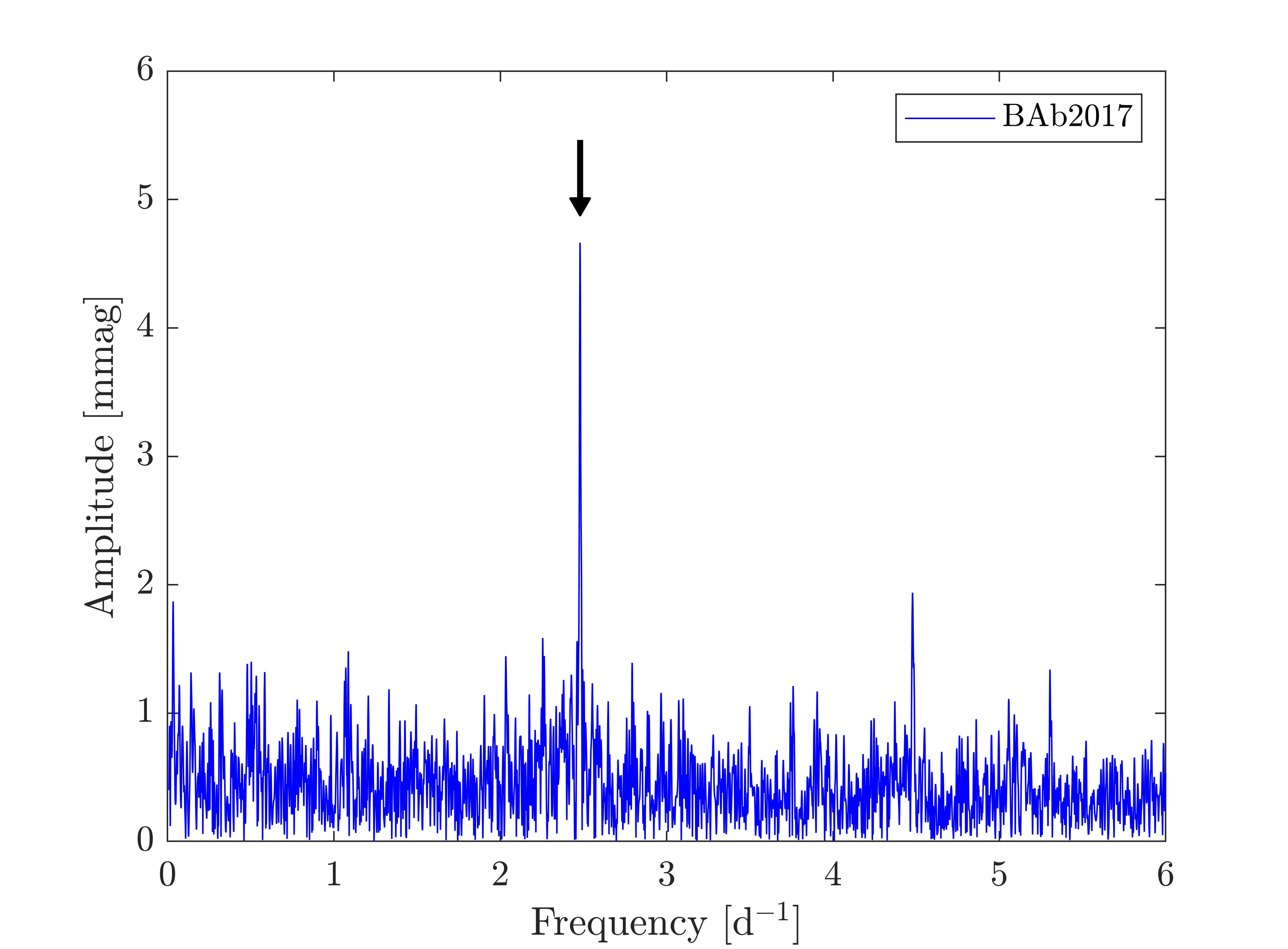}
		\caption{}
		\label{Fig:Power2017}
	\end{subfigure}
	~
\begin{subfigure}[b]{0.45\textwidth}
	\includegraphics[width=\textwidth]{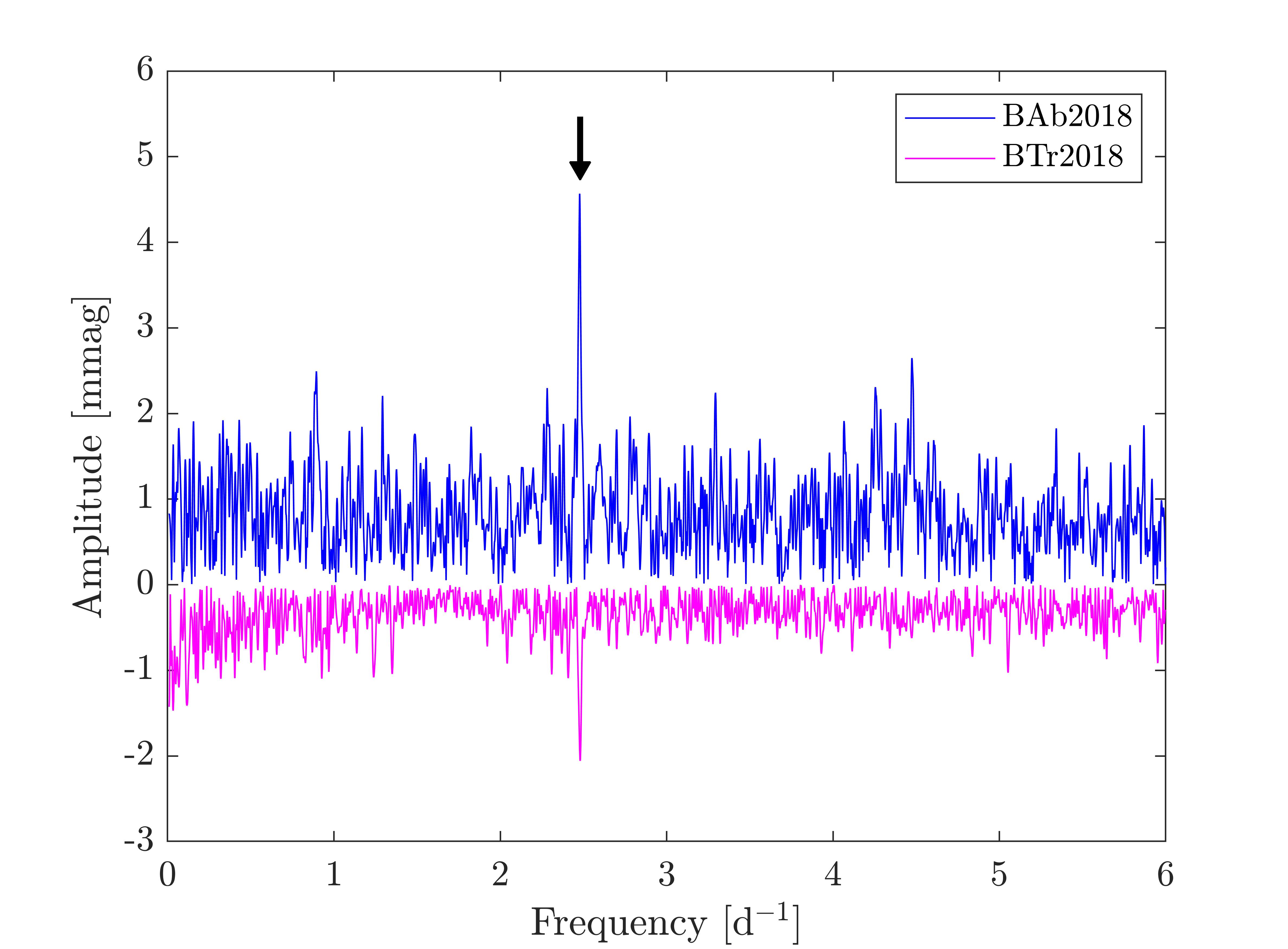}
	\caption{}
	\label{Fig:Power2018}
\end{subfigure}
\caption{Amplitude spectra for all BRITE datasets separated by satellite and season. Colours correspond to those used in Fig.\,\ref{Fig:BRITE_timeseries}. Data  in panel (a) are from 2015, (b) 2016, (c) 2017 and (d) 2019. Amplitude spectra from red-filtered satellites are flipped to negative values for clarity. The arrows mark the most prominent feature at 2.48\,d$^{-1}$.}
\label{Fig:BRITE_power}
\end{figure*}

\begin{figure}
	\centering
	\includegraphics[width=8.8cm]{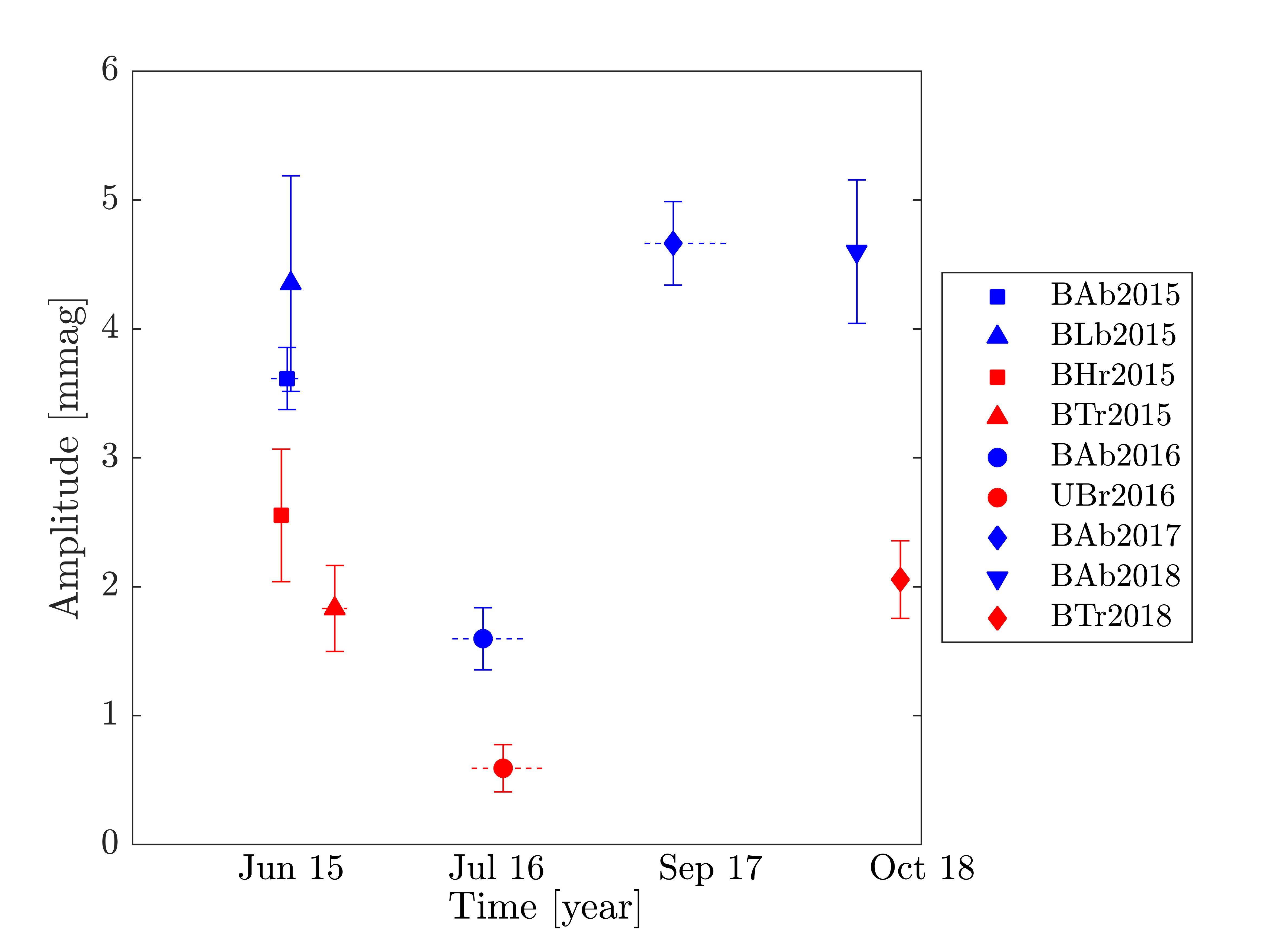}
	\caption{Amplitudes of the 2.48\,d$^{-1}$ frequency in the BRITE data. The horizontal bars indicate the time coverage of each dataset. }
	\label{fig:BRITE_amplitude}
	\end{figure}

\begin{figure}
	\centering
	\includegraphics[width=8.8cm]{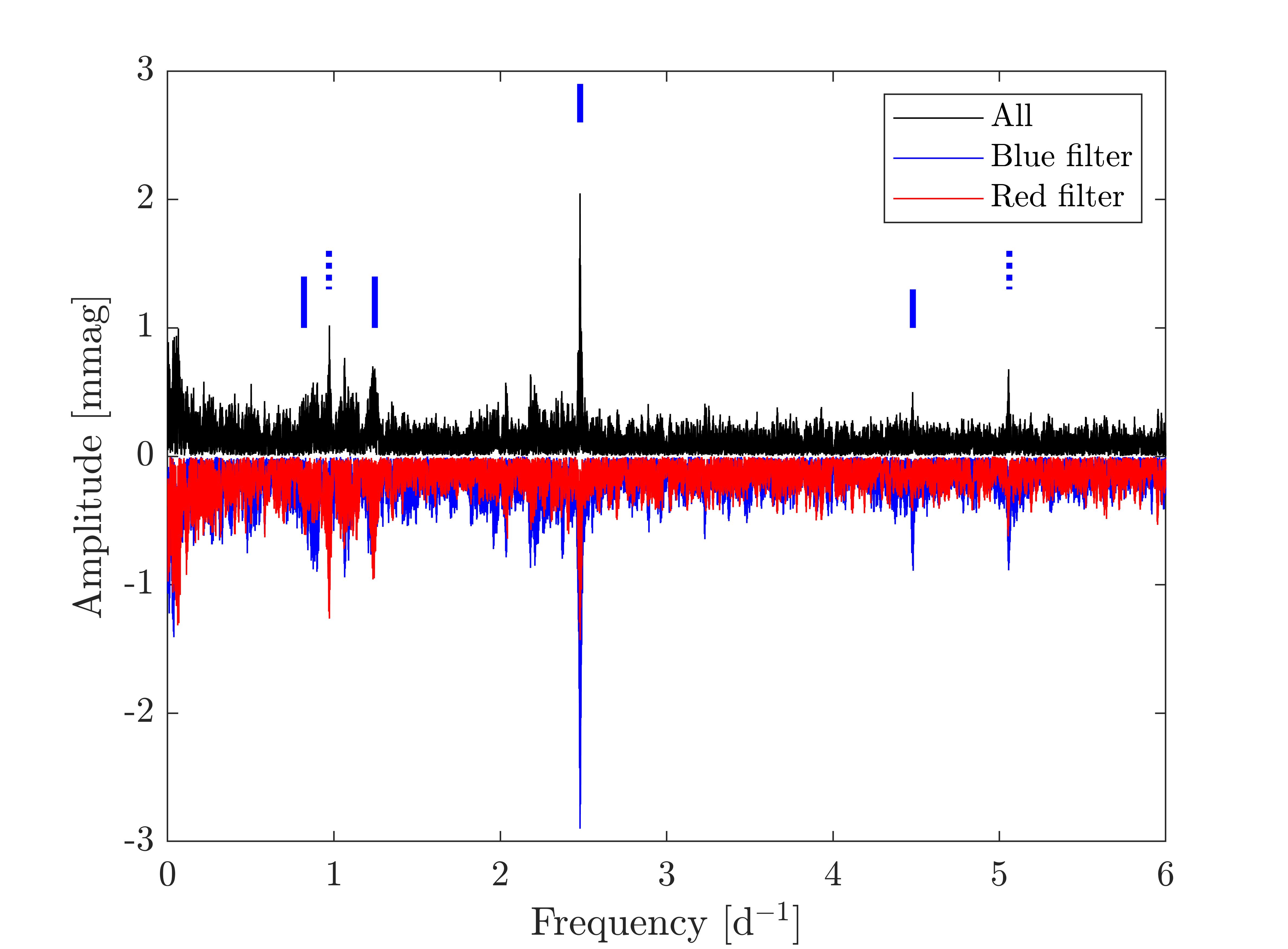}
	\caption{Amplitude spectrum of all BRITE datasets combined in black. Blue and red filter satellite data are also combined separately and plotted with corresponding colours. Blue and red spectra are flipped to negative values for clarity. Vertical full lines indicate 0.82, 1.25, 2.48 and 4.48\,d$^{-1}$ respectively. Dotted lines indicate 0.97 and 5.06\,d$^{-1}$ respectively.}
	\label{fig:BRITE_full}
	\end{figure}

The amplitude spectra  of the reduced and prewhitened BRITE data can be seen in Fig.\,\ref{Fig:BRITE_power}. 
The scale of the ordinate is not the same for all plots. 	
The most prominent feature (marked by black arrows) is at 2.48\,d$^{-1}$. This variability is present in all amplitude spectra at different amplitudes. The extracted frequencies are presented in Table\,\ref{tab:BRITE_fit} and the amplitudes are shown in Fig.\,\ref{fig:BRITE_amplitude}. 
The frequencies differ a little more than the 1$\sigma$ uncertainty. This is likely due to the short timespan in each dataset which easily enhances the effects of non-periodic signals. However, because of the short time spans of each dataset, systematic effects may become more pronounced. The amplitude spectrum of all BRITE data as well as all the blue(red) satellites combined separately can be seen in Fig.\,\ref{fig:BRITE_full}. The frequency calculated for the entire (blue+red) time series is 2.47936(2)\,d$^{-1}$. The frequency found in the SMEI data is 2.47951(4)\,d$^{-1}$ meaning the frequencies differ by 2.5$\sigma$. We emphasise that the noise in SMEI is possibly non-Gaussian which may affect the uncertainties. We attribute the difference between the frequencies to noise statistics. 

Focusing on the amplitudes in Fig.\,\ref{fig:BRITE_amplitude}, two features are noticeable: the change in amplitude over time and the difference between blue- and red-filter satellites (marked with blue and red symbols accordingly). 
In seasons when the star was observed simultaneously with a blue- and red-filter satellite, the amplitude in the blue filter was larger. Therefore, this could indicate that the signal is caused by a thermal mechanism. We have examined the ratio between the amplitudes and compared it to ratios between black body curves. However, the uncertainties on the amplitudes are too large and the number of data points is too small to make this viable. A more elaborate interpretation of a possible thermal nature of this frequency would be unjustifiable at this moment. 
The other noticeable aspect of the amplitude behaviour is the obvious fading in 2016 and recovery in 2017. Satellite BAb observed \gC during all four observing runs and thereby suggests that the changes in amplitude are not instrumental. 

We attempted to ascribe the frequency to the primary or the companion star by examining the Doppler shifts of the peaks in the BRITE amplitude spectrum which could be attributed to the orbital velocity \citep[see][ for explanation of this principle]{Hiromoto2012}. Unfortunately, the time series were too short (making the amplitude peaks too broad) for a conclusion to be drawn with confidence. 

Furthermore, we used Hubble Space Telescope\footnote{\url{www.spacetelescope.org}} (HST) data to look for the 2.48\,d$^{-1}$ signature in the UV region. HST observed \gC with the Goddard High-Resolution Spectrograph \citep{Brandt1994} between March 14th to 15th 1996 for a total of $\sim$20 hours. These data have been extensively analysed by \citet{Smith1999}. With the new knowledge of the 2.48\,d$^{-1}$ frequency we have reanalysed them to look for this signature but the length of the time series is insufficient to verify a detection.  

There is a 4-year gap between the observations with SMEI and BRITE (see Fig.\,\ref{Fig:All_timeseries}). From SMEI it was clear that the 0.82\,d$^{-1}$ frequency was decreasing in amplitude over the observing run. In agreement with this trend, the 0.82\,d$^{-1}$ frequency is not present in any of the BRITE data (individually or combined) with a detection limit of $\sim$1 mmag. 

\begin{figure}
	\centering
	\includegraphics[width=8.8cm]{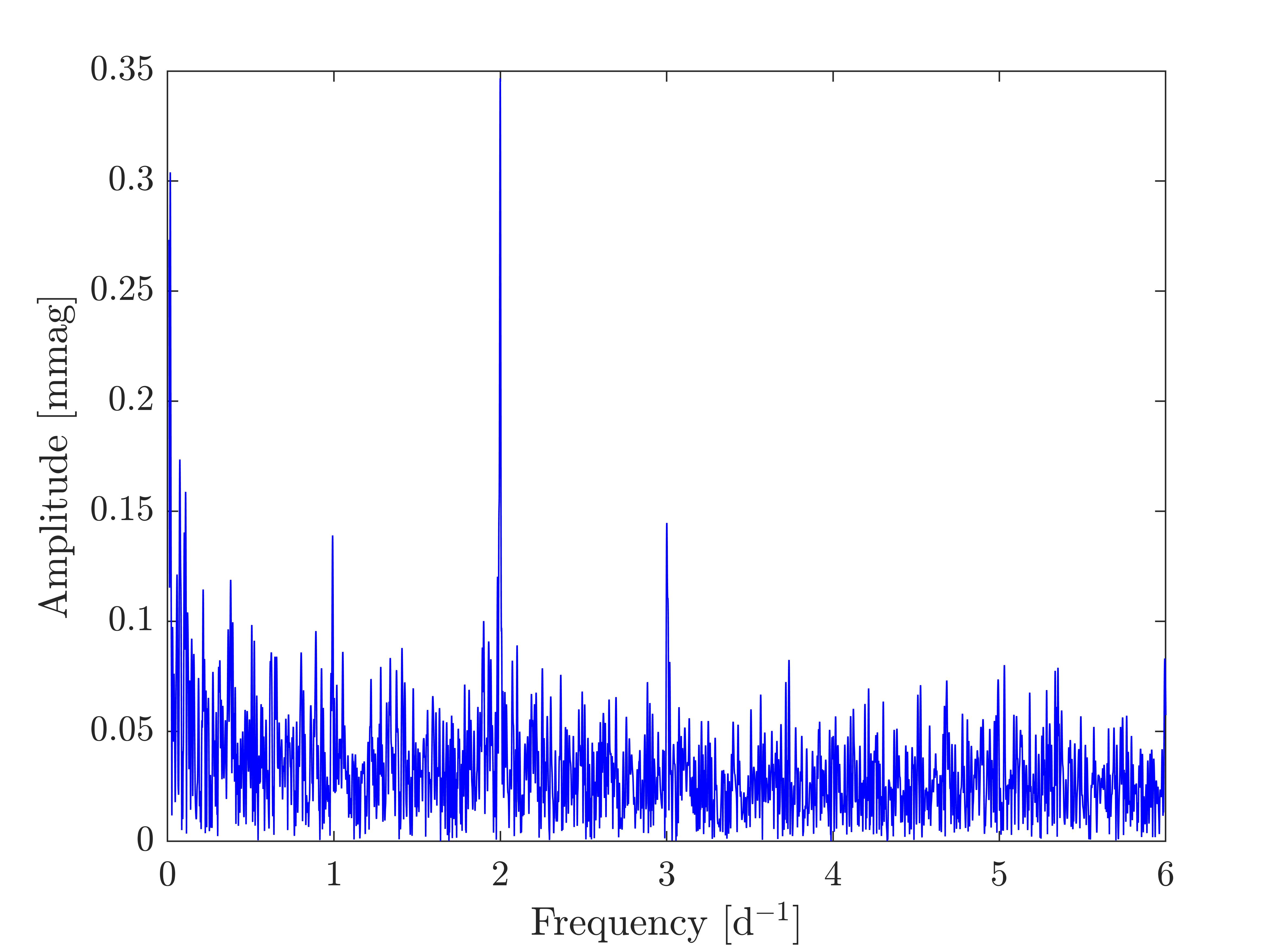}
	\caption{Window function of BRITE dataset BAb2017.}
	\label{fig:BRITE_window}
	\end{figure}

The amplitude spectra in Fig.\,\ref{Fig:BRITE_power} contain other signals that could be significant. Especially Fig.\,\ref{Fig:Power2017} and Fig.\,\ref{fig:BRITE_full} have a signal at 4.48\,d$^{-1}$. This disappears when the data are prewhitened for the 2.48\,d$^{-1}$ signal indicating that is just a daily alias. The 2\,d$^{-1}$ alias is also present in the window function and much stronger than both the 1 and 3\,d$^{-1}$ alias explaining why we do not detect the 3.48\,d$^{-1}$ (see Fig \ref{fig:BRITE_window}).
From the SMEI data we found that there is a possible signal at 1.25\,d$^{-1}$. When looking at the amplitude spectrum of the full BRITE time series (Fig.\,\ref{fig:BRITE_full}) there is an indication that it is present in the BRITE data as well. This is, however, located in a cluster of peaks and its exact frequency could not be verified with certainty. Two other peaks are marked by dashed lines in Fig.\,\ref{fig:BRITE_full}. These are located at 0.97 and 5.06\,d$^{-1}$ putting them relatively close to the daily aliases. At this point, we cannot confirm nor deny their realness.

We searched for a variation corresponding to the orbital binary period but did not find it. The BRITE datasets have all been shifted individually to zero, and this might have eliminated a coherent signal corresponding to the orbital period. The gaps in the data further complicate a possible detection.

Lastly, there are peaks of uncertain validity in all BRITE spectra. None of them appears to be consistent across multiple satellites and it is assumed that they are caused by effects intrinsic to the individual satellites (or other non-stellar effects).

	\begin{table*}
		\caption{Frequencies and amplitudes from BRITE.}
		\label{tab:BRITE_fit}
		\centering
		\tiny
	\begin{tabular}{c c c c c c c c c c c}
		\hline
			& BAb2015 & BLb2015 & BHr2015  & BTr2015  & UBr2016 & BAb2016  & BAb2017 & BAb2018 & BTr2018  & Combined \\ 
			\hline 
			Frequency [d$^{-1}$]		& 2.4802(7)  & 2.472(6)  & 2.486(2)  & 2.481(2)  & 2.478(1)  & 2.4809(7) &  2.4797(3) &2.4773(7) & 2.4799(9) & 2.47936(2)   \\ 
			Amplitude [mmag]		& 3.6(3)   & 4.5(8)  & 2.8(5)  & 1.9(3)  &     0.6(2)  & 1.7(3)   & 4.5(4) &       4.5(6) & 2.0(3) & --  \\ 
			\hline
		\end{tabular} 
	\end{table*}

	\section{Discussion}
	\label{sec:discussion}
	\subsection{Spiral structure?}
	\label{sec:discussion_spiral}
	
	The most remarkable feature seen in Fig.\,\ref{fig:GamCas_Surface2} is the blue/red-asymmetric amplitude distribution in the wings of the H$\alpha$ emission line at the orbital frequency of the companion. The simplest explanation would be a one-armed spiral which can result from the rotational distortion of the B star’s gravity field alone \citep{Okazaki1991}, i.e., without need for a companion star.  In this case, depending on the sense of rotation of the disk, the concave and the convex side of the arm would each only be seen at positive or negative radial velocities, respectively.  Because the two sides may differ not just in geometry but also in physical properties, the stellar light could be reprocessed differently and  a blue/red asymmetry of the amplitude distribution (as in Fig.\,\ref{fig:GamCas_Surface2}) would arise. The close proximity of the main feature in Fig.\,\ref{fig:GamCas_Surface2} to the orbital frequency of the companion implies that the structure is not induced by the B star's gravitational quadrupole moment only. It would be unlikely that the spiral structure's frequency coincided with the orbital frequency without them being physically linked.
	In many binary stars the gravitational effect of the companion is expected to dominate over the B star’s quadrupole moment, and a two-armed spiral would develop that is phase-locked to the companion, with one arm roughly pointing at the companion and the other one away from it \citep{Panoglou2018}.  If the two arms are identical and separated by 180 degrees in disk azimuth, any photometric signal would have to occur at twice the orbital frequency.  Accordingly, the absence of variability at double the orbital frequency implies deviations from this simplifying geometrically symmetric picture.    
	Other reasons for not observing a signal at twice the orbital frequency could be one or several of the following:  (a) the disk might not be steady (the build-up of the final spiral structure can take up to a few tens of orbits), (b) the injection of mass from the star to the disk is not isotropic, (c) the orbit is slightly eccentric, (d) orbital and disk planes are different, and/or (e) non-gravitational effects interfere, e.g. radiation from the companion ionizes one arm more than the other. The latter effect could also affect the visibility of one arm over the other, making it hard to detect even though it might be there. 
	
	Most of these effects would lead not only to variations in the optical spectra with the orbital frequency instead of its first harmonic but also to a blue/red asymmetry in the amplitude spectrum as observed in $\gamma$\,Cas (Fig.\,\ref{fig:GamCas_Surface2}).  
	Spectro-interferometry might help to downselect the number of asymmetry-causing effects such as geometry, temperature, density and velocity.

	\gC has been observed interferometrically several times in the last years. Most relevant for this discussion is the work by \cite{Berio1999} because they observed a one-armed structure in the disk using observations between 1988 and 1994. At that time it was not yet known that \gC had a companion, which could induce a two-armed spiral, but, as discussed above, the observation of only one arm does not necessarily mean that a second arm was not present. More recently, \cite{Stee2012} looked for a one-armed structure but did not find any evidence of it. They observed \gC from 2008 to 2010. Because the disk varies on timescales of a few years, it is possible that the spiral structure disappeared at some point between 1994 and 2008. This also means that, even though no spiral was observed in 2010, it can exist today, as the results from our spectroscopic analysis suggest.

	\subsection{Disk truncation?}
	\label{sec:truncation}
	The relative stability of the H$\alpha$ profiles indicates that the star has not had any large outburst in the last years. If the mass loss had ceased, the inner disk would reaccrete and the outer regions dissipate with time. The time scale for the dissipation depends on various parameters e.g. outburst rate, mass outflow rate and viscosity, but is of the order of a year or less \citep{Ghoreyshi2018}. However, the disk has been present over a number of years (according to the BeSS database at least from 2006 to 2017, see Fig.\,\ref{Fig:bindata}) and does not seem to be disappearing. The disk is therefore likely being regularly replenished but since this only happens in the inner parts of the disk, the overall structure of H$\alpha$ is roughly constant. The inner parts may also restore structure more quickly and since the observed variability is mostly at high velocities (inner parts) this could further contribute to the apparent stability of the H$\alpha$ line. 
	An explanation for this quiet nature could be linked to the companion's influence on the disk. From the flat top profile and the orbital frequency found in the disk (as opposed to its harmonics, see Sect.\,\ref{sec:BeSS}), it is evident that the companion does have a strong effect on the disk.  
	
	Outbursts are typical of Be stars \citep[e.g.][]{Bernhard2018,Labadie2017}. Probably, they are the (main) mechanism through which these stars build and sustain their disks. \cite{Panoglou2016} have also shown that the truncation by the companion can limit the mass flow across the truncation radius. It is possible that if the Be star ejects sufficient amounts of matter and if the disk cannot grow past the truncation radius, it might become thicker instead of wider. If the inner region becomes dense enough to be optically thick, the outbursts would no longer cause a brightening, meaning that the seemingly stable nature of the disk is not due to the absence of outbursts. If this is the case, it is also possible that continued growth of the disk in geometrical thickness could cause it to enter our line of sight. Shell lines can then occur due to absorption of light from the innermost parts of the disk. This could explain why in the 1930s, \gC displayed a shell phase right after large outbursts  \citep[][and references therein]{Harmanec2002} despite the 45$^\circ$ inclination \citep[][]{Quirrenbach1997} of the disk. 
	
	In general three mechanisms seem to be able to cause the line profile structure of the Be stars to change between wine-bottle structure and shell phase. One is the above-mentioned varying geometrical thickness of the disk. The second is a varying viewing aspect (i.e., azimuth) angle of the disk as is shown in simulations by \citet{Panoglou2018}.  Thirdly, if the star has a companion with a misaligned orbit it causes the disk to precess and thereby, more or less periodically, enter our line of sight. This is, for example, the case for the Be star \object{Pleione} \citep{Hirata2007}. It has not been shown that \gC's disk is (or is not) precessing. If it is not precessing, the mass build-up of the disk might be an explanation for the appearance of shell components in the  H$\alpha$ profiles.

	\subsection{Non-radial pulsation?}
	\label{sec:discussion_period}
	\begin{table}
		\centering
		\caption[Frequencies from SMEI, APT and BRITE.]{Frequencies from SMEI, APT and BRITE. APT result from  \cite{Smith2019}.}
		\label{tab:all_freq}
		\hspace*{-1cm}	\begin{tabular}{c c c c}
		\hline
			& SMEI & APT & BRITE  \\
			\hline
			Frequency [d$^{-1}$] & 0.82215(1) & 0.82247(2) & \\
						                & 1.24583(2) &            & \\
			                & 2.47951(4) &            &  2.47936(2) \\
            \hline
		\end{tabular} 
	\end{table}

	We have recovered the 0.82\,d$^{-1}$ frequency found by \cite{Smith2006} and \cite{Henry2012} using data from the SMEI satellite. The frequency was observed from 2003 to 2011, and in this time frame, it faded in amplitude until it became undetectable. We show that when the star was observed again in 2015 with BRITE, the 0.82\,d$^{-1}$ frequency was still undetectable but a new frequency of 2.48\,d$^{-1}$ now dominates.

	\cite{Smith2006} ascribed the 0.82\,d$^{-1}$ frequency to the rotation of the primary star. Their main argument for doing so was a match with expected values for the rotation rate. From radial velocity measurements and stellar interferometry, they estimate the rotation period to be 1.08-1.41 days (0.71-0.93\,d$^{-1}$) corresponding with the observed 1.21 day period (0.82\,d$^{-1}$). The other plausible contender in the explanation was NRP. 
	This was disregarded because, according to \citet{Henry2012}, there was no evidence of NRPs of the order of one day in early-type Be stars known at the time. This was despite the existence of works such as that of \citet{Rivinius2003}.
	A few years later CoRoT discovered NRPs on these timescales in many O9-B1.5-type stars \citep[e.g.][]{Degroote2009,Briquet2011}.  This was acknowledged by \cite{Henry2012}, but they argued that these studies are for slow or intermediate rotators and cannot necessarily be applied to the rapidly rotating B star in $\gamma$\,Cas. However, since then, periods of the order of one day were found in several other very early-type Be stars \citep[e.g.][]{Baade_conf,Semaan2018}. At this time it seems much more likely for Be stars to have NRPs than not.

	Although \gC is a rapid rotator, 2.48\,d$^{-1}$ interpreted as rotation frequency would lead to surface velocities larger than the break-up velocity \citep[near-critical rotation is at 0.71-0.93\,d$^{-1}$,][]{Smith2006} so that any rotational interpretation is impossible. 
	Potentially the 2.48\,d$^{-1}$ frequency could be due to the rotation of the secondary. The amplitudes (for both frequencies) are, however, unexpectedly large for a companion that is already greatly outshined by the primary in the sensitivity range of BRITE. For this reason, we assume that both frequencies are associated with the primary and that the 2.48\,d$^{-1}$ frequency is not a rotational signature.

	The ratios between the two frequencies are calculated using combinations of the values found in Table\,\ref{tab:all_freq}. When both frequencies are from SMEI the ratio is 3.01588(5), from SMEI and BRITE it is 3.01570(4) and from APT and BRITE it is 3.01453(8). Although these differ slightly they are all significantly different from 3, excluding the possibility that these are just harmonics. As mentioned, the uncertainties of the frequencies are calculated with one of the most widely used tools of the field i.e. Period04 \citep[][]{period04}.
	Although SMEI and BRITE are very different satellites, the uncertainties of the frequencies from them are of the same order of magnitude when the BRITE data are combined. On the one hand, this is due to the higher S/N of the BRITE data and, on the other one, due to the longer time baseline of the SMEI data which happen to balance out. It should be noted that the uncertainties provided here assume that the noise is normally distributed which may not be the case for SMEI. They also do not take systematics into account. Therefore, for a more conservative estimate, we combine the SMEI/SMEI and BRITE/SMEI to a ratio of 3.0158(1). This is still far from 3 and we conclude that the 0.82\,d$^{-1}$ and 2.48\,d$^{-1}$ frequencies are not related. Since they are not harmonics and not both due to rotation, the most likely origin is NRPs. 
	
    A further indication of multi-frequency NRP is the  1.25\,d$^{-1}$ frequency. Although not as prominent as the other two frequencies, the detection of this signal in two independent datasets (SMEI and BRITE) enhances its significance and supports the NRP nature of all three signals. The ratio between 2.48\,d$^{-1}$ and 1.25\,d$^{-1}$ is 1.99025(5), which is close to 2, but differs to within the nominal uncertainties. Even in the case that we greatly underestimate the uncertainties and the harmonic relations are real, either 2.48\,d$^{-1}$ is the independent frequency or both 1.25\,d$^{-1}$ and 0.82\,d$^{-1}$ are independent. This is so because, in that case, the only frequency that can be the parent of all three variations is 2.48\,d$^{-1}$. At a ratio of $\sim$1.5,  1.25\,d$^{-1}$ and 0.82\,d$^{-1}$ are not harmonics.  Therefore, if one of the two is independent, so is the other. In either case, the two higher frequencies fall into the domain of NRP frequencies observed in other Be stars but are far outside the range of rotation.  There may be further signals present in the star but more data are needed to verify this.
	
	Nearly all Be stars are observed to have NRPs \citep[e.g.][]{Rivinius2003} so it is not surprising to see them in \gC. Furthermore, the interaction of multiple NRP modes may be the driving agent of the star-to-disk mass-transfer events (outbursts) in Be stars \citep[e.g.][]{Baade2018,Baade2018_conf} for which the magnetic model of Smith and coworkers does not offer an alternative explanation.

	\section{Conclusions}
	\label{sec:conclusion}
	We have confirmed the 0.82215(1)\,d$^{-1}$ frequency  (1.21 day period) first presented by \cite{Smith2006} and \cite{Henry2012}. Using SMEI with a long time baseline, we showed that this frequency is present from 2003 to 2011 with decreasing amplitude. BRITE photometry from 2015 to 2019 supports the disappearance of this frequency with a detection limit of  $\sim$1 mmag. In BRITE data, we have found a dominant signal at a frequency of 2.47936(2)\,d$^{-1}$ (period of 0.4 days). Reanalysis of the SMEI data lets it appear possible that this 2.48\,d$^{-1}$ signal was already present at some point between 2003-2011. Since the 2.48\,d$^{-1}$ variation cannot be due to rotation, a common origin would impact the interpretation that the 1.21 day variation is the rotational signature. If they are of the same nature they would most likely be NRPs. The fact that nearly all Be stars are observed to have NRPs strengthens this explanation. If neither the 2.48\,d$^{-1}$ nor the 0.82\,d$^{-1}$ signals are NRP frequencies, \gC would be a highly unusual Be star for which no plausible explanation for these frequencies (or at least the higher frequency) exists. The possible detection of a 1.24583(4)\,d$^{-1}$ frequency in both SMEI and BRITE data would further enhance this conclusion. It was shown that at least the 0.82\,d$^{-1}$ and 1.25\,d$^{-1}$ frequencies or the 2.48\,d$^{-1}$ frequency is independent which still favours a NRP interpretation over a rotational one.

	With \gC being a very bright star, the NRP question can probably be settled through a high-cadence series of high-quality spectra. Such data could also unambiguously answer the question of whether the migrating subfeatures \citep{Yang1988} in line profiles are due to a pulsational velocity field or absorbing cloudlets. (In the case of Spica (a rapidly rotating B1 star without emission lines), \citet{Smith1985} has, in contrast to \gC, advocated NRP as the explanation of migrating subfeatures.  Apart from Spica not having the risk of contamination from circumstellar material, there does not seem to be a basic qualitative difference between these variations in $\gamma$\,Cas and Spica.)
	For such an analysis, the time series must cover at least one half-dozen rotational periods to enable a significant identification. The time span of the HST observations was too short (see Sect.\,\ref{sec:BRITE:analysis}) for this which we have confirmed with our time series analysis of these data. 
	 If at the time of such future observations the 0.82\,d$^{-1}$ variability has come back, the spectra might also identify what physical property is modulating the light. 
	
	A periodic aspect change of the spiral arms of the disk was identified in the variability of the H$\alpha$ emission profile, corresponding to the orbital frequency. Remarkably, the amplitude distribution with wavelength of this frequency displayed a blue/red asymmetry. We discussed how this could be caused by a spiral structure in the disk in phase with the companion. Due to different opacities, the concave and convex parts of the spiral could reprocess the stellar light differently. This could give rise to a different brightness when the arm is moving towards (blue) and away (red) from the observer which could produce such an asymmetry. It was, however, not possible to conclude if the spiral structure is one- or two-armed. If the spiral has two arms, the absence of a significant feature at twice the orbital frequency in the amplitude spectrum shows that any two arms are not symmetric in orbital phase and/or strength. Future observational experiments might try to address these questions more specifically.

	\begin{acknowledgements}
	We would like to thank the referee Dr. Joachim Puls for carefully reviewing the manuscript and suggesting changes that have improved the quality of the paper. 
	This work is based on CB's Master Thesis "Variability of the Former Prototype Be star and X-ray Binary $\gamma$ Cassiopeiae" (October 2018, Ludwig Maximilian University of Munich, unpublished). 
	The authors thank  the  BRITE  operations  staff for  their  untiring  efforts  to  deliver  data  of the  quality  that  enabled  this  investigation.
	The authors thank Dr. Bernard Jackson for providing the SMEI data with camera identification which made it possible to improve this work.
	This work has made use of the BeSS Database, operated at LESIA, Observatoire de Meudon, France: \url{http://basebe.obspm.fr}. 
	This research has made use of NASA’s Astrophysics Data System (ADS).
	This research has made use of the SIMBAD database, operated at CDS, Strasbourg, France.
	CB thanks the European Southern Observatory as well as the Max Planck Institute for Astrophysics for hosting her during this research. CB acknowledges  financial support from  Knud H\o jgaards Fond (grant: 16-01-2164), Dansk Tennis Fond, Nordea-fonden (grant: 01-2016-001046) and Oticon Fonden (grant: 16-3182). Funding for the Stellar Astrophysics Centre is provided by The Danish National Research Foundation (Grant agreement no.: DNRF106).
	DP acknowledges financial support from Conselho Nacional de
	Desenvolvimento Cient\'ifico e Tecnol\'ogico (Brazil) through grant
	300235/2017-8. APi acknowledges support from the NCN grant no. 2016/21/B/ST9/01126. AFJM is grateful for financial aid from NSERC (Canada) and FQRNT (Quebec). GAW also acknowledges NSERC (Canada). GH acknowledges support from the NCN grant 2015/18/A/ST9/00578. APo was responsible for image processing and automation of photometric routines for the data registered by BRITE-nanosatellite constellation, and was supported by Rector Grant no. 02/140/RGJ20/0001.
	
	\end{acknowledgements}
	
	%
	\bibliographystyle{aa.bst} 
	\bibliography{masterbib} 

\begin{thebibliography}{90}
\expandafter\ifx\csname natexlab\endcsname\relax\def\natexlab#1{#1}\fi

\bibitem[{{Abt} \& {Levy}(1978)}]{Abt&Levy1978}
{Abt}, H.~A. \& {Levy}, S.~G. 1978, \apjs, 36, 241

\bibitem[{{Baade}(1988)}]{Baade1988}
{Baade}, D. 1988, in IAU Symposium, Vol. 132, The Impact of Very High S/N
  Spectroscopy on Stellar Physics, ed. G.~{Cayrel de Strobel} \& M.~{Spite},
  217

\bibitem[{{Baade} {et~al.}(2018{\natexlab{a}}){Baade}, {Pigulski}, {Rivinius},
  {Carciofi}, {Panoglou}, {Ghoreyshi}, {Handler}, {Kuschnig}, {Moffat},
  {Pablo}, {Popowicz}, {Wade}, {Weiss}, \& {Zwintz}}]{Baade2018}
{Baade}, D., {Pigulski}, A., {Rivinius}, T., {et~al.} 2018{\natexlab{a}}, \aap,
  610, A70

\bibitem[{{Baade} {et~al.}(2017){Baade}, {Rivinius}, {Pigulski}, {Carciofi},
  {Handler}, {Kuschnig}, {Martayan}, {Mehner}, {Moffat}, {Pablo}, {Popowicz},
  {Rucinski}, {Wade}, {Weiss}, \& {Zwintz}}]{Baade_conf}
{Baade}, D., {Rivinius}, T., {Pigulski}, A., {et~al.} 2017, in Second
  BRITE-Constellation Science Conference: Small satellites - big science,
  Proceedings of the Polish Astronomical Society volume 5, ed. K.~{Zwintz} \&
  E.~{Poretti}, 196--205

\bibitem[{{Baade} {et~al.}(2018{\natexlab{b}}){Baade}, {Rivinius}, {Pigulski},
  {Panoglou}, {Carciofi}, {Handler}, {Kuschnig}, {Martayan}, {Mehner},
  {Moffat}, {Pablo}, {Popowicz}, {Rucinski}, {Wade}, {Weiss}, \&
  {Zwintz}}]{Baade2018_conf}
{Baade}, D., {Rivinius}, T., {Pigulski}, A., {et~al.} 2018{\natexlab{b}}, in
  3rd BRITE Science Conference, ed. G.~A. {Wade}, D.~{Baade}, J.~A. {Guzik}, \&
  R.~{Smolec}, Vol.~8, 69--76

\bibitem[{{Balbus}(2003)}]{Balbus2003}
{Balbus}, S.~A. 2003, \araa, 41, 555

\bibitem[{{Baldwin}(1940)}]{Baldwin1940}
{Baldwin}, R.~B. 1940, \apj, 92, 82

\bibitem[{{Berio} {et~al.}(1999){Berio}, {Stee}, {Vakili}, {Mourard},
  {Bonneau}, {Chesneau}, {Le Mignant}, {Thureau}, \& {Hirata}}]{Berio1999}
{Berio}, P., {Stee}, P., {Vakili}, F., {et~al.} 1999, \aap, 345, 203

\bibitem[{{Bernhard} {et~al.}(2018){Bernhard}, {Otero}, {H{\"u}mmerich},
  {Kaltcheva}, {Paunzen}, \& {Bohlsen}}]{Bernhard2018}
{Bernhard}, K., {Otero}, S., {H{\"u}mmerich}, S., {et~al.} 2018, \mnras, 479,
  2909

\bibitem[{{Brandt} {et~al.}(1994){Brandt}, {Heap}, {Beaver}, {Boggess},
  {Carpenter}, {Ebbets}, {Hutchings}, {Jura}, {Leckrone}, {Linsky}, {Maran},
  {Savage}, {Smith}, {Trafton}, {Walter}, {Weymann}, {Ake}, {Bruhweiler},
  {Cardelli}, {Lindler}, {Malumuth}, {Randall}, {Robinson}, {Shore}, \&
  {Wahlgren}}]{Brandt1994}
{Brandt}, J.~C., {Heap}, S.~R., {Beaver}, E.~A., {et~al.} 1994, \pasp, 106, 890

\bibitem[{{Breger} {et~al.}(1999){Breger}, {Handler}, {Garrido}, {Audard},
  {Zima}, {Papar{\'o}}, {Beichbuchner}, {Li}, {Jiang}, {Liu}, {Zhou}, {Pikall},
  {Stankov}, {Guzik}, {Sperl}, {Krzesinski}, {Ogloza}, {Pajdosz}, {Zola},
  {Thomassen}, {Solheim}, {Serkowitsch}, {Reegen}, {Rumpf}, {Schmalwieser}, \&
  {Montgomery}}]{Breger1999}
{Breger}, M., {Handler}, G., {Garrido}, R., {et~al.} 1999, \aap, 349, 225

\bibitem[{{Briquet} {et~al.}(2011){Briquet}, {Aerts}, {Baglin}, {Nieva},
  {Degroote}, {Przybilla}, {Noels}, {Schiller}, {Vu{\v c}kovi{\'c}}, {Oreiro},
  {Smolders}, {Auvergne}, {Baudin}, {Catala}, {Michel}, \&
  {Samadi}}]{Briquet2011}
{Briquet}, M., {Aerts}, C., {Baglin}, A., {et~al.} 2011, \aap, 527, A112

\bibitem[{{Buffington} {et~al.}(2007){Buffington}, {Bisi}, {Clover}, {Hick}, \&
  {Jackson}}]{Buffington2007}
{Buffington}, A., {Bisi}, M.~M., {Clover}, J.~M., {Hick}, P.~P., \& {Jackson},
  B.~V. 2007, AGU Fall Meeting Abstracts, SH33A

\bibitem[{{Carciofi} {et~al.}(2012){Carciofi}, {Bjorkman}, {Otero}, {Okazaki},
  {{\v S}tefl}, {Rivinius}, {Baade}, \& {Haubois}}]{Carciofi2012}
{Carciofi}, A.~C., {Bjorkman}, J.~E., {Otero}, S.~A., {et~al.} 2012, \apjl,
  744, L15

\bibitem[{{Collins}(1987)}]{Collins1987}
{Collins}, II, G.~W. 1987, in IAU Colloq. 92: Physics of Be Stars, ed.
  A.~{Slettebak} \& T.~P. {Snow}, 3--19

\bibitem[{{de Mink} {et~al.}(2011){de Mink}, {Langer}, \&
  {Izzard}}]{deMink2011}
{de Mink}, S.~E., {Langer}, N., \& {Izzard}, R.~G. 2011, Bulletin de la Societe
  Royale des Sciences de Liege, 80, 543

\bibitem[{{de Mink} {et~al.}(2013){de Mink}, {Langer}, {Izzard}, {Sana}, \& {de
  Koter}}]{deMink2013}
{de Mink}, S.~E., {Langer}, N., {Izzard}, R.~G., {Sana}, H., \& {de Koter}, A.
  2013, \apj, 764, 166

\bibitem[{{Degroote} {et~al.}(2009){Degroote}, {Aerts}, {Ollivier}, {Miglio},
  {Debosscher}, {Cuypers}, {Briquet}, {Montalb{\'a}n}, {Thoul}, {Noels}, {De
  Cat}, {Balaguer-N{\'u}{\~n}ez}, {Maceroni}, {Ribas}, {Auvergne}, {Baglin},
  {Deleuil}, {Weiss}, {Jorda}, {Baudin}, \& {Samadi}}]{Degroote2009}
{Degroote}, P., {Aerts}, C., {Ollivier}, M., {et~al.} 2009, \aap, 506, 471

\bibitem[{{Ekstr{\"o}m} {et~al.}(2008){Ekstr{\"o}m}, {Meynet}, {Maeder}, \&
  {Barblan}}]{Ekstrom2008}
{Ekstr{\"o}m}, S., {Meynet}, G., {Maeder}, A., \& {Barblan}, F. 2008, \aap,
  478, 467

\bibitem[{{Frandsen} {et~al.}(1995){Frandsen}, {Jones}, {Kjeldsen}, {Viskum},
  {Hjorth}, {Andersen}, \& {Thomsen}}]{Frandsen1995}
{Frandsen}, S., {Jones}, A., {Kjeldsen}, H., {et~al.} 1995, \aap, 301, 123

\bibitem[{{Fr{\'e}mat} {et~al.}(2005){Fr{\'e}mat}, {Zorec}, {Hubert}, \&
  {Floquet}}]{Fremat2005}
{Fr{\'e}mat}, Y., {Zorec}, J., {Hubert}, A.-M., \& {Floquet}, M. 2005, \aap,
  440, 305

\bibitem[{{Ghoreyshi} {et~al.}(2018){Ghoreyshi}, {Carciofi}, {R{\'{\i}}mulo},
  {Vieira}, {Faes}, {Baade}, {Bjorkman}, {Otero}, \&
  {Rivinius}}]{Ghoreyshi2018}
{Ghoreyshi}, M.~R., {Carciofi}, A.~C., {R{\'{\i}}mulo}, L.~R., {et~al.} 2018,
  \mnras, 479, 2214

\bibitem[{{Goss} {et~al.}(2011){Goss}, {Karoff}, {Chaplin}, {Elsworth}, \&
  {Stevens}}]{Goss2011}
{Goss}, K.~J.~F., {Karoff}, C., {Chaplin}, W.~J., {Elsworth}, Y., \& {Stevens},
  I.~R. 2011, \mnras, 411, 162

\bibitem[{{Granada} {et~al.}(2013){Granada}, {Ekstr{\"o}m}, {Georgy}, {Krti{\v
  c}ka}, {Owocki}, {Meynet}, \& {Maeder}}]{Granada2013}
{Granada}, A., {Ekstr{\"o}m}, S., {Georgy}, C., {et~al.} 2013, \aap, 553, A25

\bibitem[{{G{\"u}del} \& {Naz{\'e}}(2009)}]{Gudel&Naze2009}
{G{\"u}del}, M. \& {Naz{\'e}}, Y. 2009, \aapr, 17, 309

\bibitem[{{Hanuschik}(1986)}]{Hanuschik1986}
{Hanuschik}, R.~W. 1986, \aap, 166, 185

\bibitem[{{Hanuschik}(1995)}]{Hanuschik1995}
{Hanuschik}, R.~W. 1995, \aap, 295, 423

\bibitem[{{Hanuschik} {et~al.}(1996){Hanuschik}, {Hummel}, {Sutorius},
  {Dietle}, \& {Thimm}}]{Hanuschik1996}
{Hanuschik}, R.~W., {Hummel}, W., {Sutorius}, E., {Dietle}, O., \& {Thimm}, G.
  1996, \aaps, 116, 309

\bibitem[{{Harmanec}(2002)}]{Harmanec2002}
{Harmanec}, P. 2002, in Astronomical Society of the Pacific Conference Series,
  Vol. 279, Exotic Stars as Challenges to Evolution, ed. C.~A. {Tout} \&
  W.~{van Hamme}, 221

\bibitem[{{Harmanec} {et~al.}(2000){Harmanec}, {Habuda}, {{\v S}tefl},
  {Hadrava}, {Kor{\v c}{\'a}kov{\'a}}, {Koubsk{\'y}}, {Krti{\v c}ka},
  {Kub{\'a}t}, {{\v S}koda}, {{\v S}lechta}, \& {Wolf}}]{Harmanec2000}
{Harmanec}, P., {Habuda}, P., {{\v S}tefl}, S., {et~al.} 2000, \aap, 364, L85

\bibitem[{{Heard}(1938)}]{Heard1938}
{Heard}, J.~F. 1938, \jrasc, 32, 353

\bibitem[{{Henry} \& {Smith}(2012)}]{Henry2012}
{Henry}, G.~W. \& {Smith}, M.~A. 2012, \apj, 760, 10

\bibitem[{{Hick} {et~al.}(2005){Hick}, {Buffington}, \& {Jackson}}]{Hick2005}
{Hick}, P., {Buffington}, A., \& {Jackson}, B.~V. 2005, in \procspie, Vol.
  5901, Solar Physics and Space Weather Instrumentation, ed. S.~{Fineschi} \&
  R.~A. {Viereck}, 340--346

\bibitem[{{Hick} {et~al.}(2007){Hick}, {Buffington}, \& {Jackson}}]{Hick2007}
{Hick}, P., {Buffington}, A., \& {Jackson}, B.~V. 2007, 6689, 66890C

\bibitem[{{Hirata}(2007)}]{Hirata2007}
{Hirata}, R. 2007, in Astronomical Society of the Pacific Conference Series,
  Vol. 361, Active OB-Stars: Laboratories for Stellare and Circumstellar
  Physics, ed. A.~T. {Okazaki}, S.~P. {Owocki}, \& S.~{Stefl}, 267

\bibitem[{{Horne} \& {Marsh}(1986)}]{Horne1986}
{Horne}, K. \& {Marsh}, T.~R. 1986, \mnras, 218, 761

\bibitem[{{Jackson} {et~al.}(2004){Jackson}, {Buffington}, {Hick}, {Altrock},
  {Figueroa}, {Holladay}, {Johnston}, {Kahler}, {Mozer}, {Price}, {Radick},
  {Sagalyn}, {Sinclair}, {Simnett}, {Eyles}, {Cooke}, {Tappin}, {Kuchar},
  {Mizuno}, {Webb}, {Anderson}, {Keil}, {Gold}, \& {Waltham}}]{Jackson2004}
{Jackson}, B.~V., {Buffington}, A., {Hick}, P.~P., {et~al.} 2004, \solphys,
  225, 177

\bibitem[{{Jaschek} {et~al.}(1981){Jaschek}, {Slettebak}, \&
  {Jaschek}}]{Jaschek1981}
{Jaschek}, M., {Slettebak}, A., \& {Jaschek}, C. 1981, {Be star terminology.},
  Be Star Newsletter

\bibitem[{{Jernigan}(1976)}]{Jernigan1976}
{Jernigan}, J.~G. 1976, \iaucirc, 2900

\bibitem[{{Kato}(1983)}]{Kato1983}
{Kato}, S. 1983, \pasj, 35, 249

\bibitem[{{Kjeldsen}(1992)}]{Kjeldsen_phd}
{Kjeldsen}, H. 1992, PhD thesis, University of Aarhus, Denmark

\bibitem[{{Krti{\v{c}}ka} {et~al.}(2011){Krti{\v{c}}ka}, {Owocki}, \&
  {Meynet}}]{krticka2011}
{Krti{\v{c}}ka}, J., {Owocki}, S.~P., \& {Meynet}, G. 2011, \aap, 527, A84

\bibitem[{{K{\v r}{\'{\i}}{\v z}} \& {Harmanec}(1975)}]{Kriz&Harmanec1975}
{K{\v r}{\'{\i}}{\v z}}, S. \& {Harmanec}, P. 1975, Bulletin of the
  Astronomical Institutes of Czechoslovakia, 26, 65

\bibitem[{{Labadie-Bartz} {et~al.}(2017){Labadie-Bartz}, {Pepper}, {McSwain},
  {Bjorkman}, {Bjorkman}, {Lund}, {Rodriguez}, {Stassun}, {Stevens}, {James},
  {Kuhn}, {Siverd}, \& {Beatty}}]{Labadie2017}
{Labadie-Bartz}, J., {Pepper}, J., {McSwain}, M.~V., {et~al.} 2017, \aj, 153,
  252

\bibitem[{{Langer} {et~al.}(2019){Langer}, {Baade}, {Bodensteiner}, {Greiner},
  {Rivinius}, {Martayan}, \& {Borre}}]{Langer2019}
{Langer}, N., {Baade}, D., {Bodensteiner}, J., {et~al.} 2019, arXiv e-prints,
  arXiv:1911.06508

\bibitem[{{Lee} {et~al.}(1991){Lee}, {Osaki}, \& {Saio}}]{Lee1991}
{Lee}, U., {Osaki}, Y., \& {Saio}, H. 1991, \mnras, 250, 432

\bibitem[{{Lenz} \& {Breger}(2005)}]{period04}
{Lenz}, P. \& {Breger}, M. 2005, Communications in Asteroseismology, 146, 53

\bibitem[{{Meilland} {et~al.}(2012){Meilland}, {Millour}, {Kanaan}, {Stee},
  {Petrov}, {Hofmann}, {Natta}, \& {Perraut}}]{Meilland2012}
{Meilland}, A., {Millour}, F., {Kanaan}, S., {et~al.} 2012, \aap, 538, A110

\bibitem[{{Miroshnichenko} {et~al.}(2002){Miroshnichenko}, {Bjorkman}, \&
  {Krugov}}]{Miroshnichenko2002}
{Miroshnichenko}, A.~S., {Bjorkman}, K.~S., \& {Krugov}, V.~D. 2002, \pasp,
  114, 1226

\bibitem[{{Montgomery} \& {O'Donoghue}(1999)}]{Montgomery1999}
{Montgomery}, M.~H. \& {O'Donoghue}, D. 1999, Delta Scuti Star Newsletter, 13,
  28

\bibitem[{{Naz{\'e}} \& {Motch}(2018)}]{Naze2018}
{Naz{\'e}}, Y. \& {Motch}, C. 2018, \aap, 619, A148

\bibitem[{{Neiner} {et~al.}(2011){Neiner}, {de Batz}, {Cochard}, {Floquet},
  {Mekkas}, \& {Desnoux}}]{Neiner2011}
{Neiner}, C., {de Batz}, B., {Cochard}, F., {et~al.} 2011, \aj, 142, 149

\bibitem[{{Nemravov{\'a}} {et~al.}(2012){Nemravov{\'a}}, {Harmanec},
  {Koubsk{\'y}}, {Miroshnichenko}, {Yang}, {{\v S}lechta}, {Buil}, {Kor{\v
  c}{\'a}kov{\'a}}, \& {Votruba}}]{Nemravova2012}
{Nemravov{\'a}}, J., {Harmanec}, P., {Koubsk{\'y}}, P., {et~al.} 2012, \aap,
  537, A59

\bibitem[{{Okazaki}(1991)}]{Okazaki1991}
{Okazaki}, A.~T. 1991, \pasj, 43, 75

\bibitem[{{Okazaki} {et~al.}(2002){Okazaki}, {Bate}, {Ogilvie}, \&
  {Pringle}}]{Okazaki2002}
{Okazaki}, A.~T., {Bate}, M.~R., {Ogilvie}, G.~I., \& {Pringle}, J.~E. 2002,
  \mnras, 337, 967

\bibitem[{{Oudmaijer} \& {Parr}(2010)}]{Oudmaijer&Perr2010}
{Oudmaijer}, R.~D. \& {Parr}, A.~M. 2010, \mnras, 405, 2439

\bibitem[{{Owocki} {et~al.}(1994){Owocki}, {Cranmer}, \&
  {Blondin}}]{Owocki1994}
{Owocki}, S.~P., {Cranmer}, S.~R., \& {Blondin}, J.~M. 1994, \apss, 221, 455

\bibitem[{{Pablo} {et~al.}(2016){Pablo}, {Whittaker}, {Popowicz}, {Mochnacki},
  {Kuschnig}, {Grant}, {Moffat}, {Rucinski}, {Matthews},
  {Schwarzenberg-Czerny}, {Handler}, {Weiss}, {Baade}, {Wade},
  {Zoc{\l}o{\'n}ska}, {Ramiaramanantsoa}, {Unterberger}, {Zwintz}, {Pigulski},
  {Rowe}, {Koudelka}, {Orlea{\'n}ski}, {Pamyatnykh}, {Neiner}, {Wawrzaszek},
  {Marciniszyn}, {Romano}, {Wo{\'z}niak}, {Zawistowski}, \& {Zee}}]{Pablo2016}
{Pablo}, H., {Whittaker}, G.~N., {Popowicz}, A., {et~al.} 2016, \pasp, 128,
  125001

\bibitem[{{Panoglou} {et~al.}(2016){Panoglou}, {Carciofi}, {Vieira}, {Cyr},
  {Jones}, {Okazaki}, \& {Rivinius}}]{Panoglou2016}
{Panoglou}, D., {Carciofi}, A.~C., {Vieira}, R.~G., {et~al.} 2016, \mnras, 461,
  2616

\bibitem[{{Panoglou} {et~al.}(2018){Panoglou}, {Faes}, {Carciofi}, {Okazaki},
  {Baade}, {Rivinius}, \& {Borges Fernandes}}]{Panoglou2018}
{Panoglou}, D., {Faes}, D.~M., {Carciofi}, A.~C., {et~al.} 2018, \mnras, 473,
  3039

\bibitem[{{Papaloizou} {et~al.}(1992){Papaloizou}, {Savonije}, \&
  {Henrichs}}]{Papaloizou1992}
{Papaloizou}, J.~C., {Savonije}, G.~J., \& {Henrichs}, H.~F. 1992, \aap, 265,
  L45

\bibitem[{{Pigulski}(2018)}]{Pigluski2018}
{Pigulski}, A. 2018, in 3rd BRITE Science Conference, ed. G.~A. {Wade},
  D.~{Baade}, J.~A. {Guzik}, \& R.~{Smolec}, Vol.~8, 175--192

\bibitem[{{Pols} {et~al.}(1991){Pols}, {Cote}, {Waters}, \& {Heise}}]{Pols1991}
{Pols}, O.~R., {Cote}, J., {Waters}, L.~B.~F.~M., \& {Heise}, J. 1991, \aap,
  241, 419

\bibitem[{{Popowicz} {et~al.}(2017){Popowicz}, {Pigulski}, {Bernacki},
  {Kuschnig}, {Pablo}, {Ramiaramanantsoa}, {Zoc{\l}o{\'n}ska}, {Baade},
  {Handler}, {Moffat}, {Wade}, {Neiner}, {Rucinski}, {Weiss}, {Koudelka},
  {Orlea{\'n}ski}, {Schwarzenberg-Czerny}, \& {Zwintz}}]{Popowicz2017}
{Popowicz}, A., {Pigulski}, A., {Bernacki}, K., {et~al.} 2017, \aap, 605, A26

\bibitem[{{Pringle}(1981)}]{Pringle1981}
{Pringle}, J.~E. 1981, \araa, 19, 137

\bibitem[{{Puls} {et~al.}(2008){Puls}, {Vink}, \& {Najarro}}]{Puls2008}
{Puls}, J., {Vink}, J.~S., \& {Najarro}, F. 2008, \aapr, 16, 209

\bibitem[{{Quirrenbach} {et~al.}(1997){Quirrenbach}, {Bjorkman}, {Bjorkman},
  {Hummel}, {Buscher}, {Armstrong}, {Mozurkewich}, {Elias}, \&
  {Babler}}]{Quirrenbach1997}
{Quirrenbach}, A., {Bjorkman}, K.~S., {Bjorkman}, J.~E., {et~al.} 1997, \apj,
  479, 477

\bibitem[{{R{\'{\i}}mulo} {et~al.}(2018){R{\'{\i}}mulo}, {Carciofi}, {Vieira},
  {Rivinius}, {Faes}, {Figueiredo}, {Bjorkman}, {Georgy}, {Ghoreyshi}, \&
  {Soszy{\'n}ski}}]{Rimulo2018}
{R{\'{\i}}mulo}, L.~R., {Carciofi}, A.~C., {Vieira}, R.~G., {et~al.} 2018,
  \mnras, 476, 3555

\bibitem[{{Rivinius} {et~al.}(2003){Rivinius}, {Baade}, \& {{\v
  S}tefl}}]{Rivinius2003}
{Rivinius}, T., {Baade}, D., \& {{\v S}tefl}, S. 2003, \aap, 411, 229

\bibitem[{{Rivinius} {et~al.}(2013){Rivinius}, {Carciofi}, \&
  {Martayan}}]{Rivinius_R}
{Rivinius}, T., {Carciofi}, A.~C., \& {Martayan}, C. 2013, \aapr, 21, 69

\bibitem[{{Rivinius} {et~al.}(1999){Rivinius}, {{\v S}tefl}, \&
  {Baade}}]{Rivinius1999}
{Rivinius}, T., {{\v S}tefl}, S., \& {Baade}, D. 1999, \aap, 348, 831

\bibitem[{{Secchi}(1866)}]{Secchi1866}
{Secchi}, A. 1866, Astronomische Nachrichten, 68, 63

\bibitem[{{Semaan} {et~al.}(2018){Semaan}, {Hubert}, {Zorec},
  {Guti{\'e}rrez-Soto}, {Fr{\'e}mat}, {Martayan}, {Fabregat}, \&
  {Eggenberger}}]{Semaan2018}
{Semaan}, T., {Hubert}, A.~M., {Zorec}, J., {et~al.} 2018, \aap, 613, A70

\bibitem[{{Shakura} \& {Sunyaev}(1973)}]{Shakura&Sunyaev1973}
{Shakura}, N.~I. \& {Sunyaev}, R.~A. 1973, \aap, 24, 337

\bibitem[{{Shibahashi} \& {Kurtz}(2012)}]{Hiromoto2012}
{Shibahashi}, H. \& {Kurtz}, D.~W. 2012, \mnras, 422, 738

\bibitem[{{Smith}(1985)}]{Smith1985}
{Smith}, M.~A. 1985, \apj, 297, 206

\bibitem[{{Smith}(2019)}]{Smith2019}
{Smith}, M.~A. 2019, \pasp, 131, 044201

\bibitem[{{Smith} {et~al.}(2006){Smith}, {Henry}, \& {Vishniac}}]{Smith2006}
{Smith}, M.~A., {Henry}, G.~W., \& {Vishniac}, E. 2006, \apj, 647, 1375

\bibitem[{{Smith} \& {Robinson}(1999)}]{Smith1999}
{Smith}, M.~A. \& {Robinson}, R.~D. 1999, \apj, 517, 866

\bibitem[{{Smith} {et~al.}(1998){Smith}, {Robinson}, \& {Corbet}}]{Smith1998}
{Smith}, M.~A., {Robinson}, R.~D., \& {Corbet}, R.~H.~D. 1998, \apj, 503, 877

\bibitem[{{Stee} {et~al.}(2012){Stee}, {Delaa}, {Monnier}, {Meilland},
  {Perraut}, {Mourard}, {Che}, {Schaefer}, {Pedretti}, {Smith}, {Lopes de
  Oliveira}, {Motch}, {Henry}, {Richardson}, {Bjorkman}, {B{\"u}cke},
  {Pollmann}, {Zorec}, {Gies}, {ten Brummelaar}, {McAlister}, {Turner},
  {Sturmann}, {Sturmann}, \& {Ridgway}}]{Stee2012}
{Stee}, P., {Delaa}, O., {Monnier}, J.~D., {et~al.} 2012, \aap, 545, A59

\bibitem[{{Struve}(1931)}]{Struve1931}
{Struve}, O. 1931, \apj, 73, 94

\bibitem[{{Tokovinin}(1997)}]{Tokovinin1997}
{Tokovinin}, A.~A. 1997, \aaps, 124, 75

\bibitem[{{{\v S}tefl} {et~al.}(2007){{\v S}tefl}, {Okazaki}, {Rivinius}, \&
  {Baade}}]{Stefl2007}
{{\v S}tefl}, S., {Okazaki}, A.~T., {Rivinius}, T., \& {Baade}, D. 2007, in
  Astronomical Society of the Pacific Conference Series, Vol. 361, Active
  OB-Stars: Laboratories for Stellare and Circumstellar Physics, ed. A.~T.
  {Okazaki}, S.~P. {Owocki}, \& S.~{Stefl}, 274

\bibitem[{{Wade} {et~al.}(2016){Wade}, {Petit}, {Grunhut}, {Neiner}, \& {MiMeS
  Collaboration}}]{Wade2016}
{Wade}, G.~A., {Petit}, V., {Grunhut}, J.~H., {Neiner}, C., \& {MiMeS
  Collaboration}. 2016, in Astronomical Society of the Pacific Conference
  Series, Vol. 506, Bright Emissaries: Be Stars as Messengers of Star-Disk
  Physics, ed. T.~A.~A. {Sigut} \& C.~E. {Jones}, 207

\bibitem[{{Webb} {et~al.}(2006){Webb}, {Mizuno}, {Buffington}, {Cooke},
  {Eyles}, {Fry}, {Gentile}, {Hick}, {Holladay}, {Howard}, {Hewitt}, {Jackson},
  {Johnston}, {Kuchar}, {Mozer}, {Price}, {Radick}, {Simnett}, \&
  {Tappin}}]{Webb2006}
{Webb}, D.~F., {Mizuno}, D.~R., {Buffington}, A., {et~al.} 2006, Journal of
  Geophysical Research (Space Physics), 111, A12101

\bibitem[{{Weiss} {et~al.}(2014){Weiss}, {Rucinski}, {Moffat},
  {Schwarzenberg-Czerny}, {Koudelka}, {Grant}, {Zee}, {Kuschnig}, {Mochnacki},
  {Matthews}, {Orleanski}, {Pamyatnykh}, {Pigulski}, {Alves}, {Guedel},
  {Handler}, {Wade}, \& {Zwintz}}]{Weiss2014}
{Weiss}, W.~W., {Rucinski}, S.~M., {Moffat}, A.~F.~J., {et~al.} 2014, \pasp,
  126, 573

\bibitem[{{Wenger} {et~al.}(2000){Wenger}, {Ochsenbein}, {Egret}, {Dubois},
  {Bonnarel}, {Borde}, {Genova}, {Jasniewicz}, {Lalo{\"e}}, {Lesteven}, \&
  {Monier}}]{Wenger2000}
{Wenger}, M., {Ochsenbein}, F., {Egret}, D., {et~al.} 2000, \aaps, 143, 9

\bibitem[{{White} {et~al.}(1982){White}, {Swank}, {Holt}, \&
  {Parmar}}]{White1982}
{White}, N.~E., {Swank}, J.~H., {Holt}, S.~S., \& {Parmar}, A.~N. 1982, \apj,
  263, 277

\bibitem[{{Yang} {et~al.}(1988){Yang}, {Ninkov}, \& {Walker}}]{Yang1988}
{Yang}, S., {Ninkov}, Z., \& {Walker}, G.~A.~H. 1988, \pasp, 100, 233

\end{thebibliography}
	  
	%

\end{document}